\RequirePackage{lineno}
\documentclass[onecolunm,superscriptaddress,preprintnumbers,reprint,amsmath,amssymb,prc]{revtex4}  
\usepackage{graphicx}
\usepackage{dcolumn}
\usepackage{bm}
\usepackage{float}
\usepackage{color}
\usepackage[colorlinks,linkcolor=blue,urlcolor=blue,citecolor=blue]{hyperref}
\usepackage{isotope}
\usepackage[normalem]{ulem}
\renewcommand{\sout}{\bgroup \color{red} \ULdepth=-0.5ex \ULset}

\usepackage{amsmath}
\usepackage{cases}
\usepackage{amssymb}
\usepackage{eqnalign}

\begin{document}

\title{Symmetric cumulant $sc_{2,4} \left \{ 4 \right \}$ and asymmetric cumulant $ac_{2} \left \{ 3 \right \}$ from transverse momentum conservation and flow}

\author{Jia-Lin Pei}
\affiliation{Key Laboratory of Nuclear Physics and Ion-beam Application~(MOE), Institute of Modern Physics, Fudan University, Shanghai $200433$, China}
\affiliation{Shanghai Research Center for Theoretical Nuclear Physics, NSFC and Fudan University, Shanghai $200438$, China}

\author{Guo-Liang Ma}
\email{glma@fudan.edu.cn}
\affiliation{Key Laboratory of Nuclear Physics and Ion-beam Application~(MOE), Institute of Modern Physics, Fudan University, Shanghai $200433$, China}
\affiliation{Shanghai Research Center for Theoretical Nuclear Physics, NSFC and Fudan University, Shanghai $200438$, China}

\author{Adam Bzdak}
\email{bzdak@fis.agh.edu.pl}
\affiliation{AGH University of Science and Technology,\\
Faculty of Physics and Applied Computer Science,
30-059 Krak\'ow, Poland}

\begin{abstract}
Multiparticle cumulants method can be used to reveal long-range collectivity in small and large colliding systems. The four-particle symmetric cumulant $sc_{2,4} \left \{ 4 \right \}$, three-particle asymmetric cumulant $ac_{2} \left \{ 3 \right \}$, and the normalized cumulants $nsc_{2,4} \left \{ 4 \right \}$ and $nac_{2} \left \{ 3 \right \}$ from the transverse momentum conservation and flow are calculated. The interplay between the two effects is also investigated. Our results are in a good agreement with the recent ATLAS measurements of multiparticle azimuthal correlations with the subevent cumulant method, which provides insight into the origin of collective flow in small systems.
\end{abstract}
\maketitle

\section{Introduction}

High-energy nucleus-nucleus (\textit{A}+\textit{A}) collisions at the Relativistic Heavy Ion Collider (RHIC) and the Large Hadron Collider (LHC) can create an extremely dense and hot environment in which confined quarks and gluons are released into a deconfined state of matter called the quark-gluon plasma (QGP) \cite{1,2,3,4,5}. One of the most significant experimental signatures of the QGP properties is the collective flow due to its sensitivity to the dynamical evolution of the QGP, which can transfer the asymmetries in the initial geometry space into the anisotropies in the final momentum space~\cite{6,7,8,9,Wang:2013qlv}. 
The magnitude of the azimuthal anisotropy in the transverse plane of the final momentum space can be quantified in terms of the Fourier expansion coefficient, $\frac{\mathrm{d} N}{\mathrm{d} \phi } 
\propto 
1+ {\textstyle \sum_{n}} 
v_{n}\cos[n(\phi -\Psi _{n} )] $ \cite{10,11,Lan:2022rrc}, where the anisotropic flow coefficients $v_{1}$, $v_{2}$, $v_{3}$, and $v_{4}$ are directed, elliptic, triangular, and quadrangular flows, respectively. Studies of collective flow have shown that the QGP is a nearly perfect fluid with strong coupling, i.e., the ratio of the shear viscosity to the entropy density $\eta /s$ is close to the minimum value of $1/4\pi$ \cite{53,Song:2010mg,Wang:2022fwq}. In addition to the collective flow, anisotropic flow in measurements contains nonflow, which includes the short-range correlation such as jets, resonance decays, and Bose-Einstein correlation, and long-range correlation such as the transverse momentum conservation (TMC) \cite{PHOBOS:2010ekr,Zhu:2005qa,12,Borghini:2000cm,Borghini:2003ur,Dasgupta:2022psm}.

There are several methods for experimentally extracting flow coefficients, such as the event plane method requiring the estimation of reaction plane \cite{11,13}, two-particle correlation associated with ridge structure \cite{14,Ma:2014pva}, or multiparticle cumulants suppressing non-flow \cite{16,15}. The appearance of ridge structure with small azimuthal separation extending far in the longitudinal direction is considered to be direct evidence for the presence of collective flow, which was first observed in the two-particle correlation between the pseudorapidity gap $\Delta \eta$ and the azimuthal angle gap $\Delta \phi$ of the particle pairs in \textit{A}+\textit{A} collisions \cite{17,18,19,20}. This has been verified as the signature of the collective flow of the final particles in \textit{A}+\textit{A} collisions, which has been reproduced by hydrodynamic models \cite{21,22,23,24,25}. However, similar ridge structures have also been observed in small systems (e.g., \textit{p}+\textit{p}, \textit{p}+\textit{A}), posing a major challenge to previous understanding \cite{26,27,28,29,30,Zhao:2022ayk,Noronha:2024dtq},  as the applicability of hydrodynamics to small systems is controversial due to their extremely small size and short lifetime \cite{31,32,33,34,35,36}. 
Recently, many theoretical models have been employed to study anisotropic flows in small systems to understand their origins, including the final-state hydrodynamics in response to geometry asymmetries in the initial state \cite{Bozek:2011if, Bozek:2015swa, Shuryak:2013ke, Bzdak:2013zma,Qin:2013bha}, the parton escape mechanism with similar hydrodynamics \cite{Bzdak:2014dia,He:2015hfa,Lin:2015ucn,Ma:2016bbw}, and the color glass condensate (CGC) as an initial state mechanism \cite{37,38,39,40,41,42,43}. In addition, both hydrodynamic and transport models have been used to study $c_{2}\left \{ 2 \right \}$ and $c_{2}\left \{ 4 \right \}$, since the multiparticle cumulant method can suppress nonflow contributions  \cite{51,52,Lin:2021mdn,Ma:2014xfa,Ma:2016hkg}.

Previous studies have found that there is a linear relationship between the flow $v_{n}$ and the corresponding eccentricity $\varepsilon _{n}$, i.e., $v_{n}\propto  \varepsilon _{n}$ \cite{58,Han:2011iy}. 
However, it has been argued that the set of flow coefficients $\left \{ v_{n} \right \}$ and the set of eccentricities $\left \{ \varepsilon_{n} \right \}$  can be linked by a response matrix, which implies that there is a nonlinear correlation between $v_{m}$ and $v_{n}$ \cite{59,61,62}. The symmetric cumulants $sc_{n,m}\left \{ 4 \right \}  =\left \langle v_{n}^{2}v_{m}^{2} \right \rangle 
-\left \langle v_{n}^{2} \right \rangle \left \langle v_{m}^{2} \right \rangle$ can be inscribed to carry a correlation between $v_{n}$ and $v_{m}$ capable of responding to the geometrical shape eccentricities $\varepsilon _{n}$ and $\varepsilon _{m}$ of the initial state phase during the evolution of the QGP, in addition to information about the interactions of the final state \cite{44,45,46}. Moreover, $ac_{n}\left \{ 3 \right \}=\left \langle v_{n}^{2}v_{2n}\cos 2n(\Psi _{n}-\Psi _{2n} )\right \rangle$ involves not only the correlation between flow harmonics $v_{n}$ and $v_{2n}$ but also the correlation between event planes $\Psi _{n}$ and $\Psi _{2n}$. The generalized symmetric cumulants $sc_{k,l,m}\left \{ 6 \right \}$ have been proposed to explore the collectivity in large systems such as Pb+Pb collisions at LHC energies \cite{57}. The ALICE experiment measured $sc_{4,2}\left \{ 4 \right \}$ and $sc_{3,2}\left \{ 4 \right \}$ and found that there is a positive correlation between $v_{2}$ and $v_{4}$, and a negative correlation between $v_{2}$ and $v_{3}$ \cite{56}, which has shown that $sc_{n,m}\left \{ 4 \right \} $ is very sensitive to the temperature dependence of $\eta /s$ in noncentral collisions \cite{60}. To suppress the nonflow contribution to $sc_{n,m}\left \{ 4 \right \} $ and $ac_{2}\left \{ 3 \right \}$, a subevent method has been proposed, in which particles in different pseudorapidity intervals are divided into two or more subevents. The subevent method has been performed in \textit{p}+\textit{p} and \textit{p}+Pb collisions using PYTHIA and HIJING models, which shows that the subevent method can indeed suppress nonflow contributions \cite{54,55}. The recent ATLAS experimental results have demonstrated that the signal of four-particle symmetric cumulant $sc_{2,4}\left \{ 4 \right \}$ and  three-particle asymmetric cumulant $ac_{2}\left \{ 3 \right \} $ gradually decreases from the standard method to the subevent method, as a result of the effective suppression of nonflow contribution from jets \cite{50}.  

In this paper, we calculate the four-particle symmetric cumulant $sc_{2,4}\left \{ 4 \right \}$, three-particle asymmetric cumulant $ac_{2}\left \{ 3 \right \} $, and their normalized cumulants $nsc_{2,4}\left \{ 4 \right \}$ and $nac_{2}\left \{ 3 \right \} $ based on transverse momentum conservation and collective flow. Compared to the recent ATLAS experimental measurements with the subevent method, we aim to understand and explore the origin of the collectivity in small systems.

\section{$sc_{2,4} \left \{ 4 \right \}$ and $ac_{2} \left \{ 3 \right \}$ from transverse momentum conservation}

First, we summarize the calculation method of the TMC, which is assumed to be the only effect of correlations between final particles. The $k$-particle probability distribution $f(\vec{p}_{1},\dots,\vec{p}_{k})$ for the $N$-particle system with imposed transverse momentum conservation is given by \cite{Bzdak:2010fd,47,48,Xie:2022kwu} 
\begin{equation}
f(\vec{p}_{1},\dots,\vec{p}_{k})=f(\vec{p}_{1})\cdots f(\vec{p}_{k})
\frac{N}{N-k}
 \text{exp}\bigg(-\frac{(\vec{p}_{1}+\cdots+\vec{p}_{k})^{2} }{(N-k)\left \langle p^{2} \right \rangle_{F}}\bigg), 
\end{equation}
where $\left \langle p^{2} \right \rangle_{F}$ denotes the mean value of $p^{2}$ over the full space $F$,
\begin{equation}
\left \langle p^{2}\right \rangle _{F}=
\frac{\int_{F}p^{2}f(\vec{p})d^{2}\vec{p}}{\int_{F}f(\vec{p})d^{2}\vec{p}}.
\end{equation}
Our goal is to calculate the four-particle symmetric cumulant $sc_{2,4}\left \{ 4 \right \} $ and three-particle asymmetric cumulant $ac_{2}\left \{ 3 \right \} $. The four-particle symmetric cumulant and and three-particle asymmetric cumulant are defined as follows:
\begin{equation}
sc_{2,4}\left \{ 4 \right \}= \left \langle e^{i2\left ( \phi _{1}-\phi _{2}  \right )+i4\left ( \phi _{3}-\phi _{4}  \right ) }  \right \rangle- \left \langle e^{i2\left ( \phi _{1}-\phi _{2}  \right ) }  \right \rangle\left \langle e^{i4\left ( \phi _{3}-\phi _{4}  \right ) }  \right \rangle,
\label{eq3}
\end{equation} 

\begin{equation}
ac_{2}\left \{ 3 \right \}  = \left \langle e^{i2\left ( \phi _{1}+\phi _{2}-2\phi _{3}  \right )} \right \rangle.
\label{eq4}
\end{equation}

\subsection{$sc_{2,4} \left \{ 4 \right \}$}

For four particles, we have
\begin{equation}
f(\vec{p}_{1},\dots,\vec{p}_{4})=f(\vec{p}_{1})\cdots f(\vec{p}_{4})
\frac{N}{N-4}
\mathrm{exp}\bigg(-\frac{p_{1}^{2}+p_{2}^{2}+p_{3}^{2}+p_{4}^{2} }{(N-4)\left \langle p^{2} \right \rangle_{F}} \bigg)\mathrm{exp}(-\Phi ),
\end{equation}
where
\begin{equation}
\Phi =\frac{2}{(N-4)\left \langle p^{2}  \right \rangle _{F} }
\sum_{i,j=1;i< j}^{4}
p_{i}p_{j}\cos (\phi_{i}-\phi_{j} ),
\end{equation}
and $p_{i}=|\vec{p}_{i}|$.

To calculate $\left \langle e^{i2\left ( \phi _{1}-\phi _{2}  \right )+i4\left ( \phi _{3}-\phi _{4}  \right ) }  \right \rangle$ at given transverse momenta $p_{1},p_{2},p_{3},$ and $p_{4}$,
\begin{equation}
\left \langle e^{i2\left ( \phi _{1}-\phi _{2}  \right )+i4\left ( \phi _{3}-\phi _{4}  \right ) }  \right \rangle
|p_{1},p_{2},p_{3},p_{4} = 
\frac{\int_{0}^{2\pi}e^{i2\left ( \phi _{1}-\phi _{2}  \right )+i4\left ( \phi _{3}-\phi _{4}  \right ) } 
\mathrm{exp}(-\Phi )d\phi _{1}\cdots d\phi _{4} }
{\int_{0}^{2\pi}
\mathrm{exp}(-\Phi )d\phi _{1}\cdots d\phi _{4}},
\end{equation}
we expand $\mathrm{exp}(-\Phi )$ in $\Phi$. In the numerator, the first nonzero term is given by $\Phi^{6}/720$ and we neglect all higher terms. In the denominator, it is enough to take the first term, $\mathrm{exp}(-\Phi)\approx 1$, since the next terms are suppressed by the power of $1/N$. To simplify our calculation, we assume that all the transverse momentum $p_{i}$ are equal. In this case, we obtain
\begin{equation}
\left \langle e^{i2\left ( \phi _{1}-\phi _{2}  \right )+i4\left ( \phi _{3}-\phi _{4}  \right ) }  \right \rangle
|p \approx \frac{5p^{12}}{16(N-4)^{6}\left \langle p^{2} \right \rangle_{F}^{6}}. 
\end{equation}
Performing analogous calculations we obtain
\begin{equation}
\left \langle e^{i2\left ( \phi _{1}-\phi _{2}  \right ) }  \right \rangle
|p \approx
\frac{p^{4}}{2(N-2)^{2}\left \langle p^{2} \right \rangle_{F}^{2}},
\end{equation}
and
\begin{equation}
\left \langle e^{i4\left ( \phi _{3}-\phi _{4}  \right ) }  \right \rangle
|p \approx
\frac{p^{8}}{24(N-2)^{4}\left \langle p^{2} \right \rangle_{F}^{4}}.
\end{equation}
Using Eq.~\eqref{eq3} we find
\begin{equation}
sc_{2,4} \left \{ 4 \right \}\approx 
\frac{5p^{12}}{16(N-4)^{6}\left \langle p^{2} \right \rangle_{F}^{6}}
-
\frac{p^{12}}{48(N-2)^{6}\left \langle p^{2} \right \rangle_{F}^{6}}.
\label{eq11}
\end{equation}

\subsection{$ac_{2} \left \{ 3 \right \}$}

For three particles, we have
\begin{equation}
f(\vec{p}_{1},\dots,\vec{p}_{3})=f(\vec{p}_{1})\cdots f(\vec{p}_{3})
\frac{N}{N-3}
 \mathrm{exp}\bigg(-\frac{p_{1}^{2}+p_{2}^{2}+p_{3}^{2} }{(N-3)\left \langle p^{2} \right \rangle_{F}} \bigg)\mathrm{exp}(-\Phi ),
\end{equation}
where
\begin{equation}
\Phi =\frac{2}{(N-3)\left \langle p^{2}  \right \rangle _{F} }
\sum_{i,j=1;i< j}^{3}
p_{i}p_{j}\cos (\phi_{i}-\phi_{j} ).
\end{equation}
Using Eq.~\eqref{eq4} we find
\begin{equation}
ac_{2} \left \{ 3 \right \}\approx 
\frac{p^{8}}{4(N-3)^{2}\left \langle p^{2} \right \rangle_{F}^{4}}.
\label{eq14}
\end{equation}

\section{$sc_{2,4} \left \{ 4 \right \}$ and $ac_{2} \left \{ 3 \right \}$ from transverse momentum conservation and flow}
Next, we calculate the contribution of the TMC and the collective flow to the four-particle symmetric cumulant $sc_{2,4} \left \{ 4 \right \}$ and three-particle asymmetric cumulant $ac_{2} \left \{ 3 \right \}$. The particle emission azimuthal angle distribution measured with respect to the reaction plane is characterized by a Fourier expansion,
 \begin{equation}
 f\left ( p ,\phi \right ) =\frac{g\left ( p \right ) }{2\pi}\left ( 1+\sum_{n}2v_{n} \left (  p\right ) \cos \left [ n\left ( \phi -\Psi _{n}  \right )  \right ]   \right ),
 \end{equation}
where $v_{n}$ and $\Psi _{n}$ denote the $nth$-order flow coefficient and the reaction plane angle. In our calculations we consider $v_{2}$, $v_{3}$, and $v_{4}$ only.

\subsection{$sc_{2,4} \left \{ 4 \right \}$}
The four-particle probability distribution with TMC can be written as \cite{48}
\begin{equation}
f_{4} \left ( p_{1},\phi _{1}  ,\dots,p_{4},\phi _{4}  \right )  =f\left ( p_{1},\phi _{1}   \right )\cdots f\left ( p_{4},\phi _{4}   \right )\frac{N}{N- 4}
\mathrm{exp}\left ( - \frac{\left ( p_{1,x}+  \dots +p_{4,x}    \right )^{2}  }{2\left (  N-4\right )\left \langle p_{x}^{2}     \right   \rangle _{F}  }  - \frac{\left ( p_{1,y}+  \dots +p_{4,y}    \right )^{2}  }{2\left (  N-4\right )\left \langle p_{y}^{2}     \right   \rangle _{F}  }\right ),
\end{equation}
where
\begin{equation}
p_{x} = p\cos \left (  \phi\right ),
\hspace{0.5cm}p_{y} = p\sin \left (  \phi\right ),
\end{equation}
\begin{equation}
\left \langle p_{x}^{2}   \right \rangle_{F}  = \frac{1}{2} \left \langle p^{2}  \right \rangle _{F}\left ( 1+  v_{2F}   \right ),
\hspace{0.5cm}\left \langle p_{y}^{2}   \right \rangle_{F}  = \frac{1}{2} \left \langle p^{2}  \right \rangle _{F}\left ( 1-  v_{2F}   \right ),
\end{equation}
\begin{equation}
v_{2F}=   \frac{\int_{F} v_{2}\left ( p \right )g(p)p^{2}d^{2}p    }{\int_{F} g(p)p^{2}d^{2}p}.
\end{equation}

Using
\begin{equation}
\left \langle e^{i2\left ( \phi _{1}-\phi _{2}  \right )+i4\left ( \phi _{3}-\phi _{4}  \right ) }  \right \rangle\mid p_{1},p_{2},p_{3},p_{4} =\frac{\int_{0}^{2\pi } e^{i2\left ( \phi _{1}-\phi _{2}  \right )+i4\left ( \phi _{3}-\phi _{4}  \right ) }f_{4} \left ( p_{1},\phi _{1}  ,\dots,p_{4},\phi _{4}  \right )d\phi _{1} \dots d\phi _{4} }{\int_{0}^{2\pi } f_{4} \left ( p_{1},\phi _{1}  ,\dots,p_{4},\phi _{4}  \right )d\phi _{1} \dots d\phi _{4}},
\end{equation}
\begin{equation}
p_{1} =p_{2} =p_{3} =p_{4} =p,
\end{equation}
and including all the terms up to the one containing the pure TMC effect, $e^{X} \approx 1+X  +\frac{X^{2} }{2} +\frac{X^{3} }{3!} +\frac{X^{4} }{4!} +\frac{X^{5} }{5!} +\frac{X^{6} }{6!}$,
we obtain
\begin{equation}
\left \langle e^{i2\left ( \phi _{1}-\phi _{2}  \right )+i4\left ( \phi _{3}-\phi _{4}  \right ) }  \right \rangle\mid p\approx A_{0}+A_{1}Y_{A}+\frac{1}{2} A_{2}Y_{A}^{2}+\frac{1}{6} A_{3}Y_{A}^{3}+\frac{1}{24} A_{4}Y_{A}^{4}+\frac{1}{120} A_{5}Y_{A}^{5}+\frac{1}{720} A_{6}Y_{A}^{6}
,
\label{eq22}
\end{equation}
where
\begin{equation}
Y _{A}=-\frac{p^{2} }{(N-4) \left \langle p^{2}  \right \rangle _{F}(1-v_{2F}^{2} )},
\end{equation}
and
\begin{align*}
A_{0} =\hspace{0.1cm}&v_{2}^{2} v_{4}^{2}, \notag \\
A_{1} =\hspace{0.1cm}&v_{2}^{2} v_{3}^{2} + 4 v_{2}^{2} v_{4}^{2} + v_{3}^{2} v_{4}^{2} - v_{2} v_{2F} v_{4}^{2} \cos(2 \Psi_{2}) + 2 v_{2} v_{3}^{2} v_{4} \cos(2 \Psi_{2} - 6 \Psi_{3} + 4 \Psi_{4})
,\notag \\
A_{2} =\hspace{0.1cm}&v_{2}^{4} + 16 v_{2}^{2} v_{3}^{2} + 4 v_{3}^{4} + v_{4}^{2} + 30 v_{2}^{2} v_{4}^{2} + \frac{{v_{2F}^{2} v_{4}^{2}}}{2} + 16 v_{3}^{2} v_{4}^{2} + v_{4}^{4} - 6 v_{2} v_{2F} v_{3}^{2} \cos(2 \Psi_{2}) - 20 v_{2} v_{2F} v_{4}^{2} \cos(2 \Psi_{2}) \notag \\ &+ 2 v_{2}^{2} v_{4} \cos(4 \Psi_{2} - 4 \Psi_{4}) - 6 v_{2F} v_{3}^{2} v_{4} \cos(6 \Psi_{3} - 4 \Psi_{4}) + 24 v_{2} v_{3}^{2} v_{4} \cos(2 \Psi_{2} - 6 \Psi_{3} + 4 \Psi_{4})
,\notag\\
A_{3} =\hspace{0.1cm}&24 v_{2}^{4} + 9 v_{3}^{2} + 234 v_{2}^{2} v_{3}^{2} + \frac{{27 v_{2F}^{2} v_{3}^{2}}}{2} + 72 v_{3}^{4} + 24 v_{4}^{2} + 304 v_{2}^{2} v_{4}^{2} + 36 v_{2F}^{2} v_{4}^{2} + 234 v_{3}^{2} v_{4}^{2} + 24 v_{4}^{4} - 21 v_{2}^{3} v_{2F} \cos(2 \Psi_{2}) \notag \\ &- 210 v_{2} v_{2F} v_{3}^{2} \cos(2 \Psi_{2}) - 369 v_{2} v_{2F} v_{4}^{2} \cos(2 \Psi_{2}) - 21 v_{2} v_{2F} v_{4} \cos(2 \Psi_{2} - 4 \Psi_{4}) + 48 v_{2}^{2} v_{4} \cos(4 \Psi_{2} - 4 \Psi_{4}) \notag \\ &- 168 v_{2F} v_{3}^{2} v_{4} \cos(6 \Psi_{3} - 4 \Psi_{4}) + 300 v_{2} v_{3}^{2} v_{4} \cos(2 \Psi_{2} - 6 \Psi_{3} + 4 \Psi_{4})
,\notag\\
A_{4} =\hspace{0.1cm}&49 v_{2}^{2} + 436 v_{2}^{4} + 147 v_{2}^{2} v_{2F}^{2} + 264 v_{3}^{2} + 3328 v_{2}^{2} v_{3}^{2} + 792 v_{2F}^{2} v_{3}^{2} + 1056 v_{3}^{4} + 448 v_{4}^{2} + 3625 v_{2}^{2} v_{4}^{2} + 1344 v_{2F}^{2} v_{4}^{2} \notag \\ &+ 3328 v_{3}^{2} v_{4}^{2} + 436 v_{4}^{4} - 872 v_{2}^{3} v_{2F} \cos(2 \Psi_{2}) - 5004 v_{2} v_{2F} v_{3}^{2} \cos(2 \Psi_{2}) - 6728 v_{2} v_{2F} v_{4}^{2} \cos(2 \Psi_{2}) \notag \\ &+ 45 v_{2}^{2} v_{2F}^{2} \cos(4 \Psi_{2}) - 872 v_{2} v_{2F} v_{4} \cos(2 \Psi_{2} - 4 \Psi_{4}) + 898 v_{2}^{2} v_{4} \cos(4 \Psi_{2} - 4 \Psi_{4}) - 3492 v_{2F} v_{3}^{2} v_{4} \cos(6 \Psi_{3} - 4 \Psi_{4}) \notag \\ &+ 45 v_{2F}^{2} v_{4} \cos(4 \Psi_{4}) + 3920 v_{2} v_{3}^{2} v_{4} \cos(2 \Psi_{2} - 6 \Psi_{3} + 4 \Psi_{4})
,\notag\\
\end{align*}

\begin{align}
A_{5} =\hspace{0.1cm}&1820 v_{2}^{2} + 7120 v_{2}^{4} + 9100 v_{2}^{2} v_{2F}^{2} + 5545 v_{3}^{2} + 47025 v_{2}^{2} v_{3}^{2} + 27725 v_{2F}^{2} v_{3}^{2} + 14800 v_{3}^{4} + 7680 v_{4}^{2} + 47004 v_{2}^{2} v_{4}^{2} \notag \\ &+ 38400 v_{2F}^{2} v_{4}^{2} + 47001 v_{3}^{2} v_{4}^{2} + 7120 v_{4}^{4} - 525 v_{2} v_{2F} \cos(2 \Psi_{2}) - 23855 v_{2}^{3} v_{2F} \cos(2 \Psi_{2}) - \frac{1575}{2} v_{2} v_{2F}^{3} \cos(2 \Psi_{2}) \notag \\ &- 102960 v_{2} v_{2F} v_{3}^{2} \cos(2 \Psi_{2}) - 120970 v_{2} v_{2F} v_{4}^{2} \cos(2 \Psi_{2}) + 2940 v_{2}^{2} v_{2F}^{2} \cos(4 \Psi_{2}) - 23895 v_{2} v_{2F} v_{4} \cos(2 \Psi_{2} - 4 \Psi_{4}) \notag \\ &+ 15480 v_{2}^{2} v_{4} \cos(4 \Psi_{2} - 4 \Psi_{4}) - 65430 v_{2F} v_{3}^{2} v_{4} \cos(6 \Psi_{3} - 4 \Psi_{4}) + 2940 v_{2F}^{2} v_{4} \cos(4 \Psi_{4}) \notag \\ &+ 52730 v_{2} v_{3}^{2} v_{4} \cos(2 \Psi_{2} - 6 \Psi_{3} + 4 \Psi_{4})
,\notag\\
A_{6} =\hspace{0.1cm}&225 + 45396 v_{2}^{2} + 110941 v_{2}^{4} + \frac{{3375 v_{2F}^{2}}}{2} + 340470 v_{2}^{2} v_{2F}^{2} + \frac{{10125 v_{2F}^{4}}}{8} + 102780 v_{3}^{2} \notag \\ &+ 666216 v_{2}^{2} v_{3}^{2} + 770850 v_{2F}^{2} v_{3}^{2} + 205956 v_{3}^{4} + 126720 v_{4}^{2}+637704v_{2}^{2}v_{4}^{2}+950400v_{2F}^{2}v_{4}^{2} + 664800 v_{3}^{2} v_{4}^{2} \notag \\ &+ 110686 v_{4}^{4} - 28140 v_{2} v_{2F} \cos(2 \Psi_{2}) - 545892 v_{2}^{3} v_{2F} \cos(2 \Psi_{2}) - 70350 v_{2} v_{2F}^{3} \cos(2 \Psi_{2}) - 1970532 v_{2} v_{2F} v_{3}^{2} \cos(2 \Psi_{2}) \notag \\ &- 2142936 v_{2} v_{2F} v_{4}^{2} \cos(2 \Psi_{2}) + 115605 v_{2}^{2} v_{2F}^{2} \cos(4 \Psi_{2}) - 548580 v_{2} v_{2F} v_{4} \cos(2 \Psi_{2} - 4 \Psi_{4}) \notag \\ &+ 257412 v_{2}^{2} v_{4} \cos(4 \Psi_{2} - 4 \Psi_{4}) - 1169892 v_{2F} v_{3}^{2} v_{4} \cos(6 \Psi_{3} - 4 \Psi_{4}) + 113610 v_{2F}^{2} v_{4} \cos(4 \Psi_{4}) \notag \\ &+ 724092 v_{2} v_{3}^{2} v_{4} \cos(2 \Psi_{2} - 6 \Psi_{3} + 4 \Psi_{4})
.
\label{eq24}
\end{align}

Note that the above items are approximate results, where we kept the terms up to $v_{n}^4$. The full results and the ratios relative to the full results are shown in Eq.~(\ref{eqA1}) and the left panel of Fig.~\ref{ratio} in the Appendix, respectively.

Similarly, we have
\begin{align}
\left \langle e^{i2\left ( \phi _{1}-\phi _{2}  \right ) }  \right \rangle
|p \approx  B_{0}+B_{1}Y_{B}+\frac{1}{2} B_{2}Y_{B}^{2},
\label{eq25}
\end{align}
where
\begin{equation}
Y_{B}=-\frac{p^{2} }{(N-2) \left \langle p^{2}  \right \rangle _{F}(1-v_{2F}^{2} )}
\end{equation}
and
\begin{align}
B_{0} =\hspace{0.1cm}&v_{2}^{2},\notag\\
B_{1} =\hspace{0.1cm}&2 v_{2}^{2} + v_{3}^{2} - v_{2} v_{2F} \cos(2 \Psi_{2}) - v_{2} v_{2F} v_{4} \cos(2 \Psi_{2} - 4 \Psi_{4}),\notag\\
B_{2} =\hspace{0.1cm}&1 + 6 v_{2}^{2} + \frac{{v_{2F}^{2}}}{2} + 3 v_{2}^{2} v_{2F}^{2} + 4 v_{3}^{2} + 2 v_{2F}^{2} v_{3}^{2} + v_{4}^{2} + \frac{{v_{2F}^{2} v_{4}^{2}}}{2} - 8 v_{2} v_{2F} \cos(2 \Psi_{2}) + \frac{{1}}{{2}} v_{2}^{2} v_{2F}^{2} \cos(4 \Psi_{2}) \notag\\ &- 8 v_{2} v_{2F} v_{4} \cos(2 \Psi_{2} - 4 \Psi_{4}) + 3 v_{2F}^{2} v_{4} \cos(4 \Psi_{4})
.\notag\\
\label{eq27}
\end{align}
Finally
\begin{equation}
\left \langle e^{i4\left ( \phi _{3}-\phi _{4}  \right ) }  \right \rangle
|p \approx 
C_{0}+C_{1}Y_{C}+\frac{1}{2} C_{2}Y_{C}^{2}+\frac{1}{6} C_{3}Y_{C}^{3}+\frac{1}{24} C_{4}Y_{C}^{4},
\label{eq28}
\end{equation}
where
\begin{equation}
Y_{C}=-\frac{p^{2} }{(N-2) \left \langle p^{2}  \right \rangle _{F}(1-v_{2F}^{2} )}
\end{equation}
and
\begin{align*}
C_{0} =\hspace{0.1cm}&v_{4}^{2},\notag\\
C_{1} =\hspace{0.1cm}&v_{3}^{2}+2v_{4}^{2}-v_{2} v_{2F} v_{4} \cos(2 \Psi_{2} - 4 \Psi_{4}),\notag\\
C_{2} =\hspace{0.1cm}&v_{2}^{2} + \frac{v_{2}^{2} v_{2F}^{2}}{2} + 4 v_{3}^{2} + 2 v_{2F}^{2} v_{3}^{2} + 6 v_{4}^{2} + 3 v_{2F}^{2} v_{4}^{2} - 8 v_{2} v_{2F} v_{4} \cos(2 \Psi_{2} - 4 \Psi_{4}) + \frac{1}{2} v_{2F}^{2} v_{4} \cos(4 \Psi_{4}),\notag\\
C_{3} =\hspace{0.1cm}&6 v_{2}^{2} + 9 v_{2}^{2} v_{2F}^{2} + 15 v_{3}^{2} + \frac{45 v_{2F}^{2} v_{3}^{2}}{2} + 20 v_{4}^{2} + 30 v_{2F}^{2} v_{4}^{2} - 3 v_{2} v_{2F} \cos(2 \Psi_{2}) - \frac{3}{4} v_{2} v_{2F}^{3} \cos(2 \Psi_{2}) \notag\\ &- 45 v_{2} v_{2F} v_{4} \cos(2 \Psi_{2} - 4 \Psi_{4}) - \frac{45}{4} v_{2} v_{2F}^{3} v_{4} \cos(2 \Psi_{2} - 4 \Psi_{4}) + 9 v_{2F}^{2} v_{4} \cos(4 \Psi_{4}) - \frac{1}{4} v_{2} v_{2F}^{3} v_{4} \cos(2 \Psi_{2} + 4 \Psi_{4}),\notag\\
\end{align*}

\begin{align}
C_{4} =\hspace{0.1cm} & 1 + 28 v_{2}^{2} + 3 v_{2F}^{2} + 84 v_{2}^{2} v_{2F}^{2} + \frac{3 v_{2F}^{4}}{8} + \frac{21 v_{2}^{2} v_{2F}^{4}}{2} + 56 v_{3}^{2} + 168 v_{2F}^{2} v_{3}^{2} + 21 v_{2F}^{4} v_{3}^{2} + 70 v_{4}^{2} + 210 v_{2F}^{2} v_{4}^{2} \notag\\ &+ \frac{105 v_{2F}^{4} v_{4}^{2}}{4} 
 - 32 v_{2} v_{2F} \cos(2 \Psi_{2}) - 24 v_{2} v_{2F}^{3} \cos(2 \Psi_{2}) + 3 v_{2}^{2} v_{2F}^{2} \cos(4 \Psi_{2}) + \frac{1}{2} v_{2}^{2} v_{2F}^{4} \cos(4 \Psi_{2}) - v_{2F}^{3} v_{3}^{2} \cos(6 \Psi_{3}) 
 \notag\\ &- 224 v_{2} v_{2F} v_{4} \cos(2 \Psi_{2} - 4 \Psi_{4}) - 168 v_{2} v_{2F}^{3} v_{4} \cos(2 \Psi_{2} - 4 \Psi_{4}) + 84 v_{2F}^{2} v_{4} \cos(4 \Psi_{4}) + 14 v_{2F}^{4} v_{4} \cos(4 \Psi_{4}) 
 \notag\\ &+ \frac{1}{8} v_{2F}^{4} v_{4}^{2} \cos(8 \Psi_{4}) - 8 v_{2} v_{2F}^{3} v_{4} \cos(2 \Psi_{2} + 4 \Psi_{4}).
\label{eq30}
\end{align}

\subsection{$ac_{2} \left \{ 3 \right \}$}

The three-particle probability distribution with TMC can be written as
\begin{equation}
f_{3} \left ( p_{1},\phi _{1}  ,\dots,p_{3},\phi _{3}  \right )  =f\left ( p_{1},\phi _{1}   \right )\cdots f\left ( p_{3},\phi _{3}   \right )\frac{N}{N-3}
\mathrm{exp}\left ( - \frac{\left ( p_{1,x}+  \dots +p_{3,x}    \right )^{2}  }{2\left (  N-3\right )\left \langle p_{x}^{2}     \right   \rangle _{F}  }  - \frac{\left ( p_{1,y}+  \dots +p_{3,y}    \right )^{2}  }{2\left (  N-3\right )\left \langle p_{y}^{2}     \right   \rangle _{F}  }\right ).
\end{equation}
We have
\begin{align}
\label{eq32}
\left \langle e^{i2\left ( \phi _{1}+\phi _{2}  -2\phi _{3}\right )} \right \rangle\mid p &=\frac{\int_{0}^{2\pi } e^{i2\left ( \phi _{1}+\phi _{2}  -2\phi _{3}\right )}f_{3} \left ( p_{1},\phi _{1}  ,\dots,p_{3},\phi _{3}  \right )d\phi _{1} \dots d\phi _{3} }{\int_{0}^{2\pi } f_{3} \left ( p_{1},\phi _{1}  ,\dots,p_{3},\phi _{3}  \right )d\phi _{1} \dots d\phi _{3}}\notag\\
&\approx 
D_{0}+D_{1}Y_{D}+\frac{1}{2} D_{2}Y_{D}^{2}+\frac{1}{6} D_{3}Y_{D}^{3}+\frac{1}{24} D_{4}Y_{D}^{4},
\end{align}
where
\begin{equation}
Y_{D}=-\frac{p^{2} }{(N-3) \left \langle p^{2}  \right \rangle _{F}(1-v_{2F}^{2} )}
\end{equation}
and
\begin{align}
D_{0} = \hspace{0.1cm}&v_{2}^{2}v_{4}\cos(4\Psi_{2} -4\Psi_{4} ),\notag\\
D_{1} = \hspace{0.1cm}&- v_{2}v_{2F}v_{4}\cos(2\Psi_{2}-4\Psi_{4})
+3 v_{2}^{2}v_{4}\cos(4\Psi_{2} -4\Psi_{4} ),\notag\\
D_{2} = \hspace{0.1cm}&2 v_{2}^{2}+4v_{3}^{2}+2 v_{4}^{2}-14 v_2 v_{2F} v_4 \cos(2 \Psi_2 - 4 \Psi_4)+15v_{2}^{2}v_{4}\cos(4\Psi_{2} -4\Psi_{4} )+\frac{3}{2}v_{2F}^{2}v_{4}\cos(4\Psi _{4} )  ,\notag\\
D_{3} = \hspace{0.1cm}&30v_{2}^{2}+48v_{3}^{2}+30v_{4}^{2}-12v_{2}v_{2F}\cos(2\Psi_{2})
-177 v_2 v_{2F} v_4 \cos(2 \Psi_2 - 4 \Psi_4)+93v_{2}^{2}v_{4}\cos(4\Psi_{2} -4\Psi_{4} )+\frac{75}{2}v_{2F}^{2}v_{4}\cos(4\Psi _{4} )  ,\notag\\
D_{4} = \hspace{0.1cm}&6+322v_{2}^{2}+456v_{3}^{2}+320v_{4}^{2}-272v_{2}v_{2F}\cos(2\Psi_{2})
-2004 v_2 v_{2F} v_4 \cos(2 \Psi_2 - 4 \Psi_4)+651v_{2}^{2}v_{4}\cos(4\Psi_{2} -4\Psi_{4} )\notag\\
&+645v_{2F}^{2}v_{4}\cos(4\Psi _{4} ).
\label{eq34}
\end{align}

Note that the above items are approximate results, where we kept the terms up to $v_{n}^3$. The full results and the ratios relative to the full results are shown in Eq.~(\ref{eqA2}) and the right panel of Fig.~\ref{ratio} in the Appendix, respectively.

\subsection{$nsc_{2,4} \left \{ 4 \right \}$ and $nac_{2} \left \{ 3 \right \}$}
The normalized cumulants are defined as follows:
\begin{align}
nsc_{2,4}\left \{ 4 \right \}&=\frac{sc_{2,4}\left \{ 4 \right \}}{v_{2}\left \{ 2 \right \}^{2} v_{4}\left \{ 2 \right \}^{2}  },\\
nac_{2}\left \{ 3 \right \} &=\frac{ac_{2}\left \{ 3 \right \}}{v_{2}\left \{ 2 \right \}^{2} \sqrt{v_{4}\left \{ 2 \right \}^{2}}   },
\end{align}
where $v_{2} \left \{ 2 \right \}$ and $v_{4} \left \{ 2 \right \}$ are from Eqs.~\eqref{eq25} and ~\eqref{eq28}. The normalized cumulants only reflect the strength of the correlation between $v_2$ and $v_4$, whereas the unnormalized cumulants have contributions from both the correlations between the two different flow harmonics and the individual harmonics.

\section{Results}

\begin{figure}[H]
\centering
\includegraphics[scale=0.4]{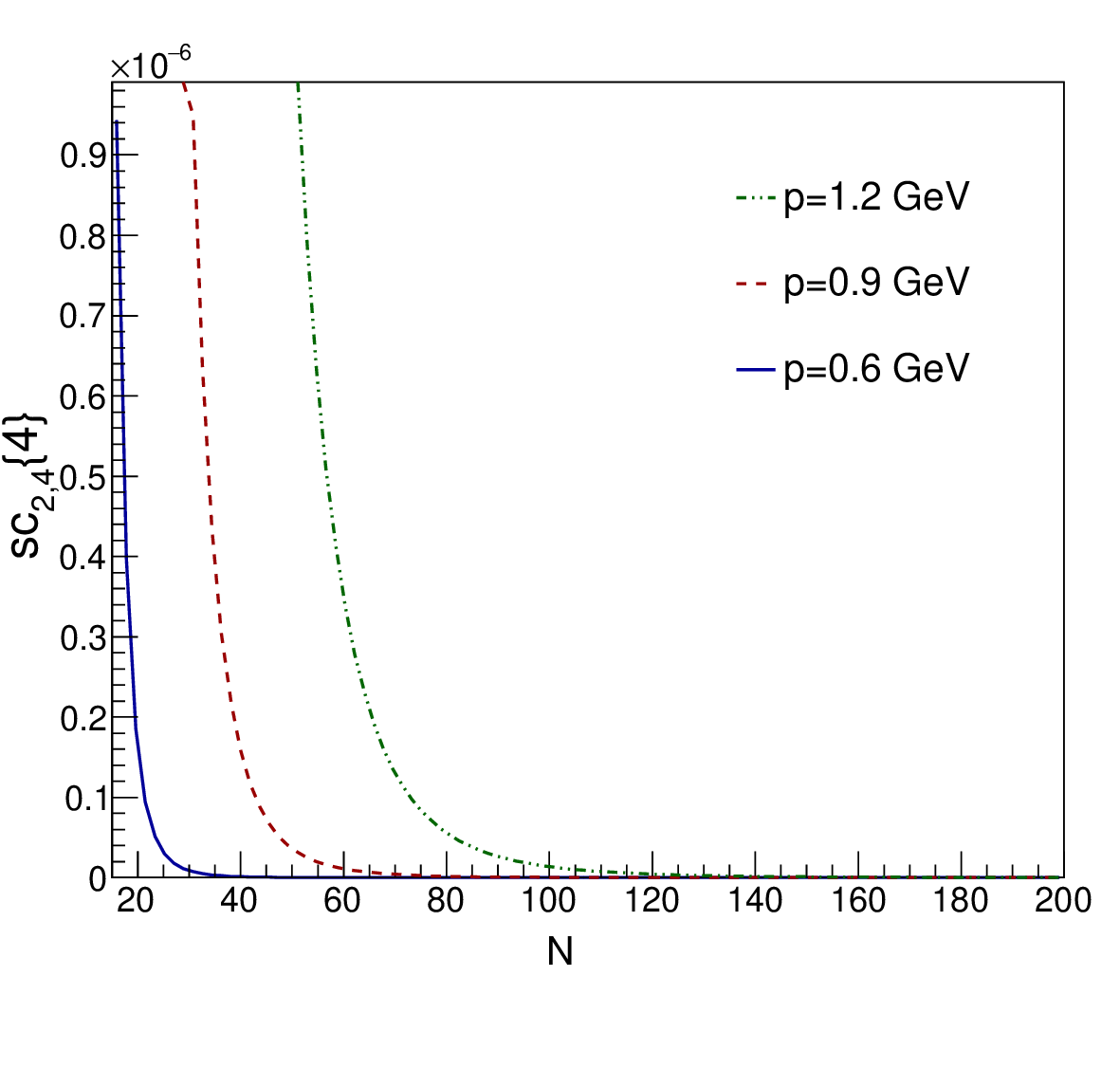}
\includegraphics[scale=0.4]{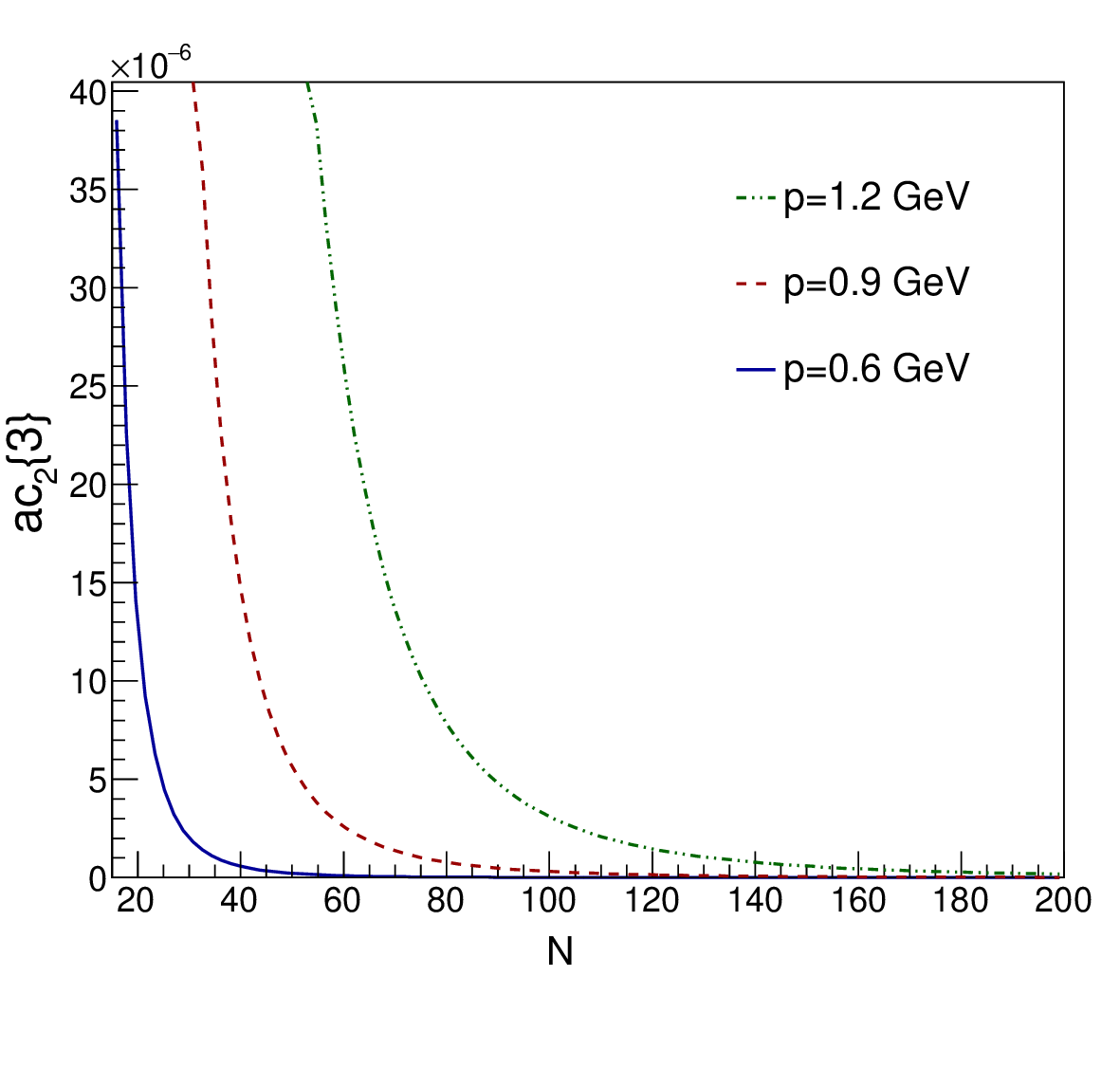}
\caption{The four-particle symmetric cumulants $sc_{2,4} \left \{ 4 \right \}$ and three-particle asymmetric cumulants $ac_{2} \left \{ 3 \right \}$ from transverse momentum conservation only as a function of the number of particles $N$ for various values of transverse momenta $p$.}
\label{purescac}
\end{figure}

Based on Eqs.~\eqref{eq11} and ~\eqref{eq14}, we present the four-particle symmetric cumulants $sc_{2,4} \left \{ 4 \right \}$ and three-particle asymmetric cumulants $ac_{2} \left \{ 3 \right \}$ from transverse momentum conservation only as a function of the number of particles $N$ for various values of transverse momenta $p=0.6, 0.9, 1.2$ GeV in Fig.~\ref{purescac}. In our calculation, $\left \langle p^{2}  \right \rangle _{F}$= 0.25 $\rm GeV^{2}$. 
It can be seen that the values of $sc_{2,4} \left \{ 4 \right \}$ and $ac_{2} \left \{ 3 \right \}$ from transverse momentum conservation decrease and tend to zero as $N$ increases, and that $sc_{2,4} \left \{ 4 \right \}$ and $ac_{2} \left \{ 3 \right \}$ also increase with transverse momenta $p$, which are consistent with the trends found in Refs. \cite{54,55} using the PYTHIA model. This is a manifestation of the property of the TMC that it is more effective at smaller $N$ and rather negligible at larger $N$. 

\begin{figure}[H]
\centering
\includegraphics[scale=0.4]{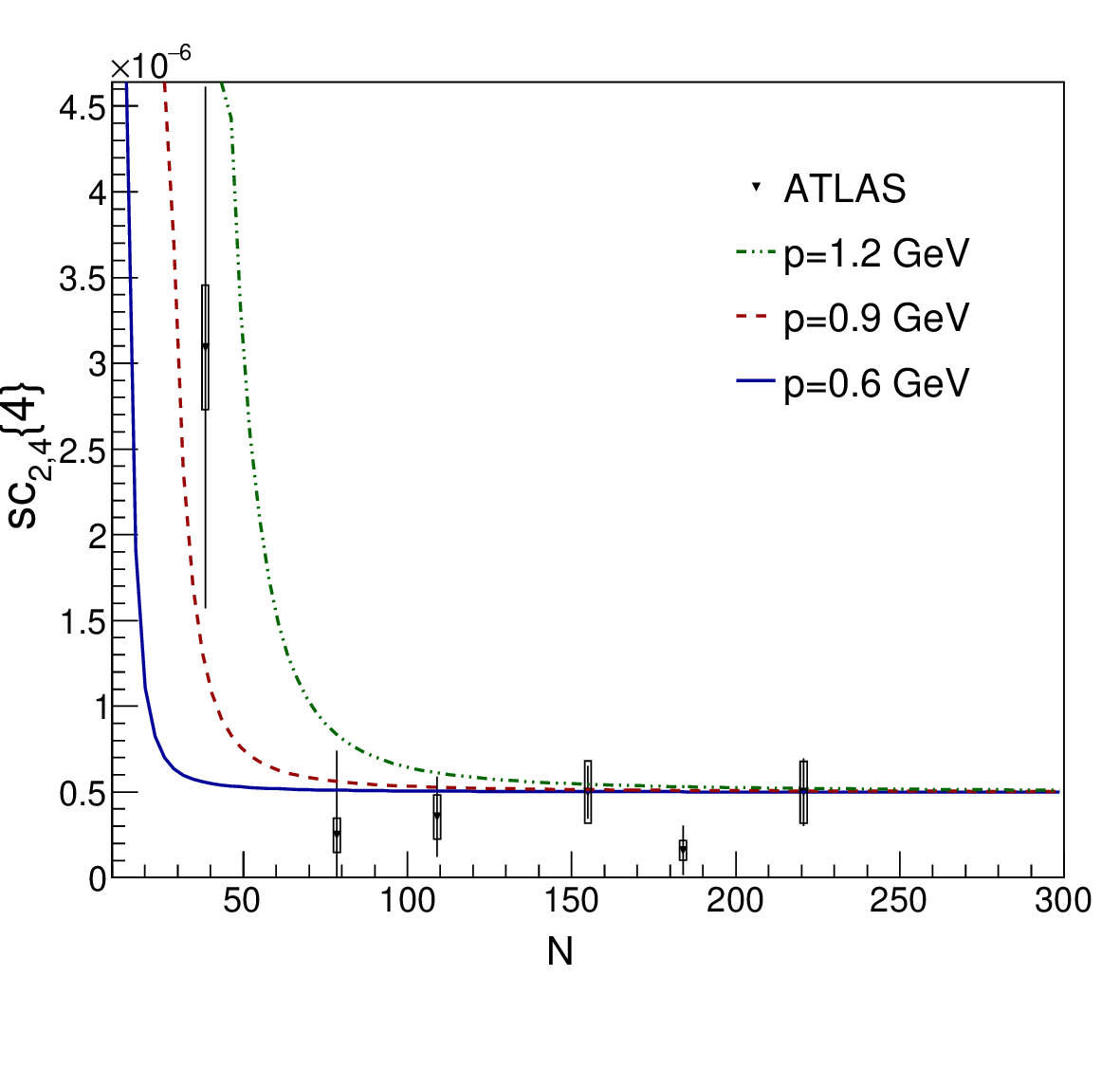}
\includegraphics[scale=0.4]{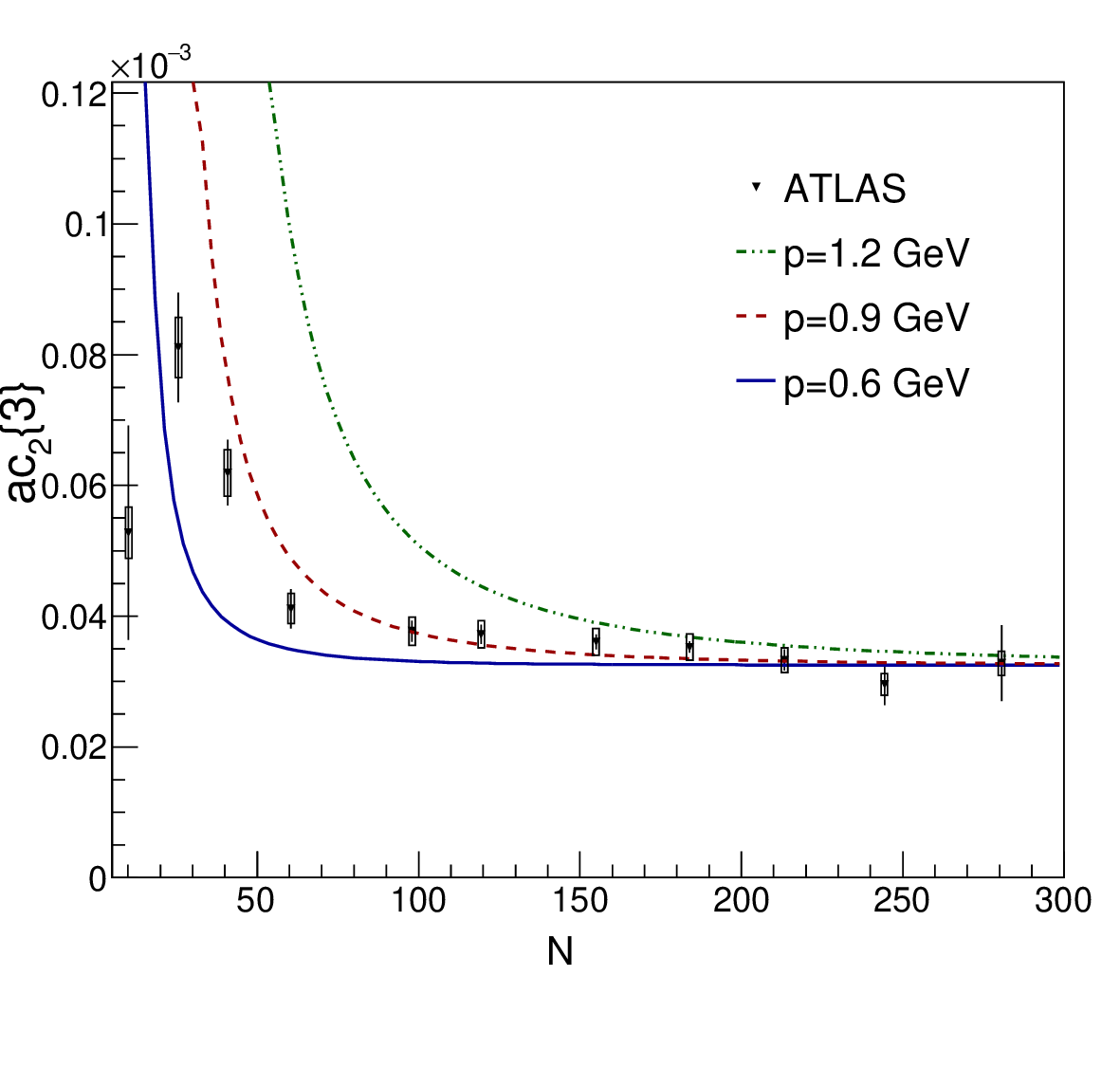}
\caption{ $sc_{2,4} \left \{ 4 \right \}$ and $ac_{2} \left \{ 3 \right \}$ from transverse momentum conservation and flow as a function of the number of particles $N$ for various values of transverse momenta $p$. The ATLAS data for $0.3< p_{T} < 3$ GeV in p+p collisions at 13 TeV using the four-subevent cumulant method or three-subevent cumulant method are shown for comparisons, where the error bars and boxes represent the statistical and systematic uncertainties, respectively \cite{50}.}
\label{scactotal}
\end{figure}

According to Eqs.~\eqref{eq22}, ~\eqref{eq25}, ~\eqref{eq28}, and~\eqref{eq32}, Fig.~\ref{scactotal} shows $sc_{2,4} \left \{ 4 \right \}$ and $ac_{2} \left \{ 3 \right \}$ from transverse momentum conservation and flow as a function of the number of particles $N$ for various values of transverse momenta $p=0.6, 0.9, 1.2$ GeV. In our calculation, we set $v_{2}= 0.08$, $v_{3}= 0.0175$, $v_{4}= 0.08^{2}$, $\left \langle p^{2}  \right \rangle _{F}= 0.5^{2}$, $v_{2F}= 0.025$, $\Psi _{2} = 0$, $\cos(4(\Psi _{4}-\Psi _{2} ))=0.8$, and $\cos(2\Psi _{2}-6\Psi _{3}+4\Psi _{4} )=-0.15$. The values of the correlations among different combinations of event planes here are from Ref. \cite{49}. We observe that both $sc_{2,4} \left \{ 4 \right \}$ and $ac_{2} \left \{ 3 \right \}$ decrease with the increase of multiplicity and their magnitudes are consistent with the data. Note that since the multiplicity $N$ refers to the number of particles under the influence of the TMC rather than the number of experimentally detected charged particles, we multiply the experimental number of charged particles by 1.5 to obtain the total number of particles $N$ in the experimental data in all figures. In comparison with Fig.~\ref{purescac}, for larger $N$, $sc_{2,4} \left \{ 4 \right \}$ and $ac_{2} \left \{ 3 \right \}$ do not converge to zero, which is caused by the existence of flow due to hydrodynamics. When $N$ is relatively small, $sc_{2,4} \left \{ 4 \right \}$ and $ac_{2} \left \{ 3 \right \}$ increase with increasing momentum $p$, whereas when $N$ is large, $sc_{2,4} \left \{ 4 \right \}$ and $ac_{2} \left \{ 3 \right \}$ hardly change with momentum. 

\begin{figure}[H]
\centering
\includegraphics[scale=0.4]{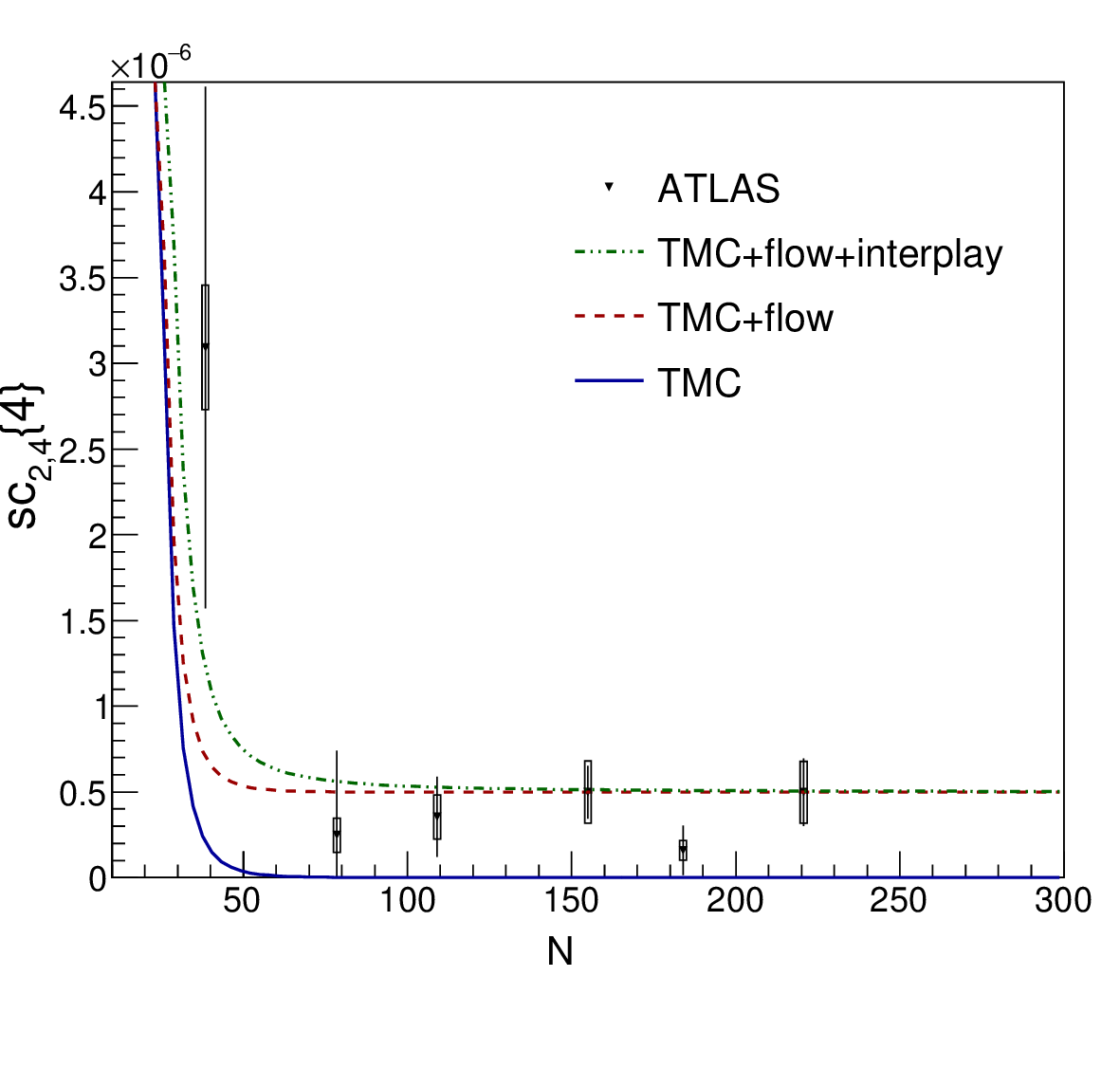}
\includegraphics[scale=0.4]{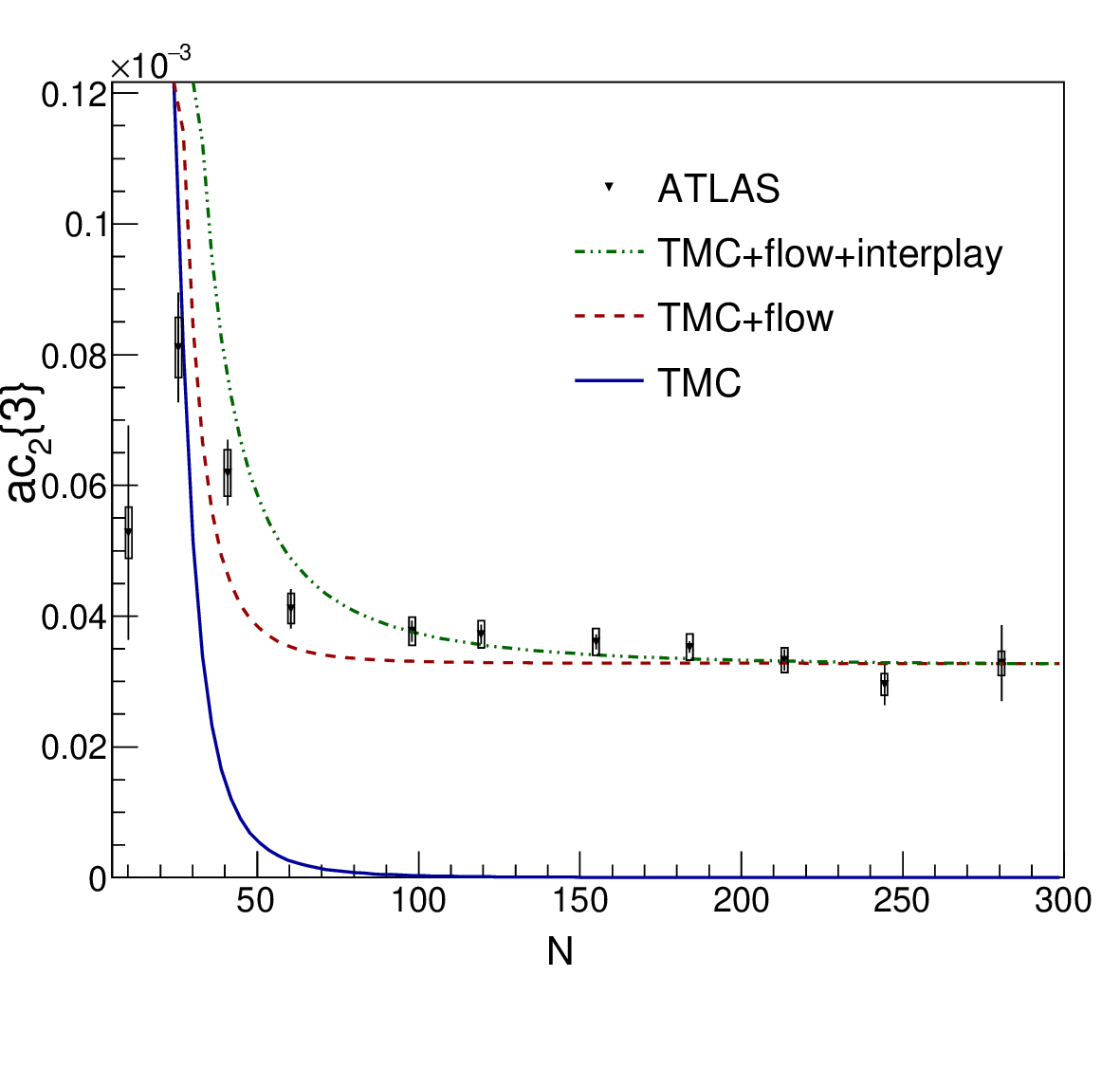}
\caption{$sc_{2,4} \left \{ 4 \right \}$ and $ac_{2} \left \{ 3 \right \}$ from the TMC, the TMC and collective flow, and plus interplay as a function of the number of particles $N$ for momentum $p=0.9$ GeV. The ATLAS data for $0.3< p_{T} < 3$ GeV in p+p collisions at 13 TeV using the four-subevent cumulant method or three-subevent cumulant method are shown for comparisons, where the error bars and boxes represent the statistical and systematic uncertainties, respectively \cite{50}.}
\label{partscac}
\end{figure}

To expound how the TMC and collective flow affect $sc_{2,4} \left \{ 4 \right \}$ and $ac_{2} \left \{ 3 \right \}$, Fig.~\ref{partscac} presents the respective contributions from the TMC only (denoted as ``TMC''), the TMC and collective flow (denoted as ``TMC+flow''), and plus interplay (denoted as ``TMC+flow+interplay'') for $p=0.9$ GeV. Here ``TMC'' refers to the terms that depend only on $N$ and $p$, ``flow" refers to the terms that depend only on $v_{n}$ and $\Psi _{n}$, and ``interplay" refers to terms that depend on both $N$, $p$, $v_{n}$, and $\Psi _{n}$ in Eqs.~\eqref{eq22}, ~\eqref{eq25}, ~\eqref{eq28}, and~\eqref{eq32}. In Fig.~\ref{partscac}, ``TMC+flow'' means the sum of ``TMC'' and ``flow'', and ``TMC+flow+interplay'' means the combination of all three of the above. We see that collective flow makes the curve higher and the contribution from interplay is present when $N$ is small, but almost negligible when $N$ is large. It can be understood as when $N$ is small, the TMC dominates, and when $N$ is large, the contribution from collective flow  becomes significant. 

\begin{figure}[H]
\centering
\includegraphics[scale=0.4]{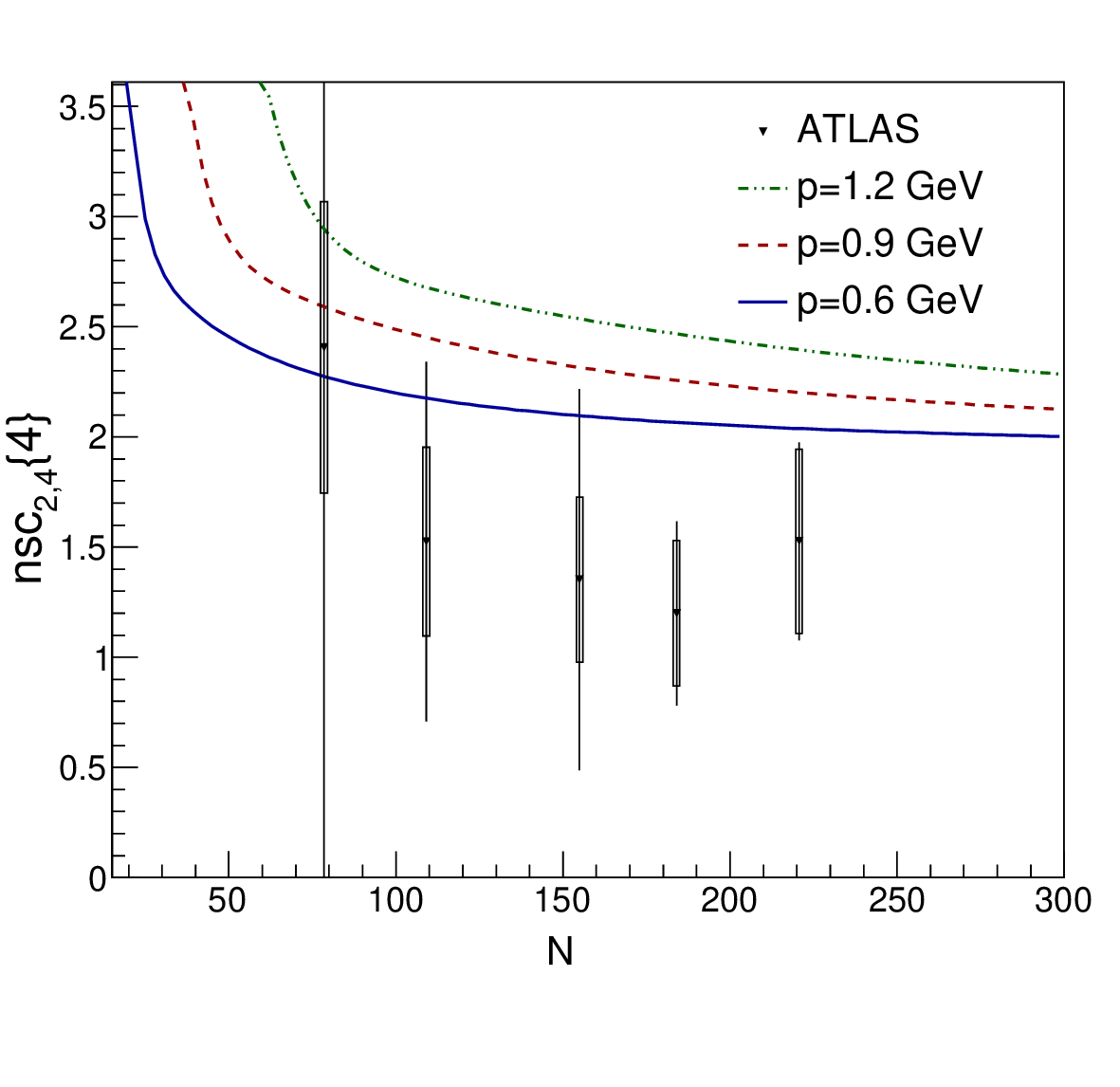}
\includegraphics[scale=0.4]{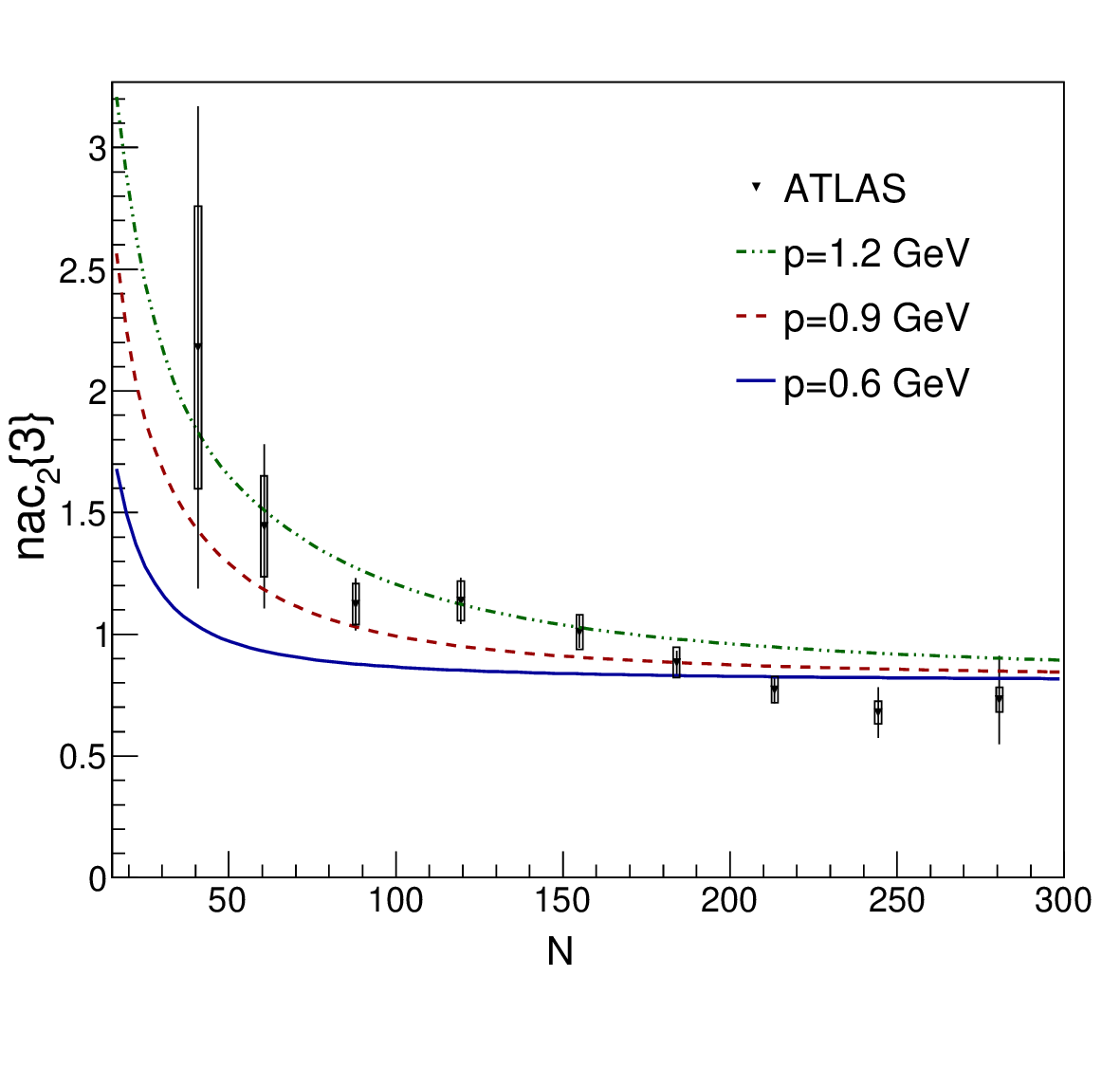}
\caption{$nsc_{2,4} \left \{ 4 \right \}$ and $nac_{2} \left \{ 3 \right \}$ from the TMC and flow as a function of the number of particles $N$ for various values of transverse momenta $p$. The ATLAS data for $0.3< p_{T} < 3$ GeV in p+p collisions at 13 TeV using the three-subevent cumulant method are shown for comparisons, where the error bars and boxes represent the statistical and systematic uncertainties, respectively \cite{50}.}
\label{nscac}
\end{figure}

Figure~\ref{nscac} shows that the normalized cumulants $nsc_{2,4} \left \{ 4 \right \}$ and $nac_{2} \left \{ 3 \right \}$ from the TMC and flow decrease with the increase of multiplicity, which can basically describe the experimental data. Since Fig.~\ref{scactotal} has shown that $sc_{2,4} \left \{ 4 \right \}$ and $ac_{2} \left \{ 3 \right \}$ can describe the ATLAS data, it indicates that our results on two-particle $v_{2} \left \{ 2 \right \}$ and $v_{4} \left \{ 2 \right \}$ should also be consistent with the experimental data. In the left plot of Fig.~\ref{nscac}, because the TMC contribution is small when $N$ is large, our results close to 2.0 for $nsc_{2,4} \left \{ 4 \right \}$ at very large $N$ reflect the correlation between $v_{2} \left \{ 2 \right \}$ and $v_{4} \left \{ 2 \right \}$ produced by hydrodynamics. In the right plot of Fig.~\ref{nscac}, we see that $nac_{2} \left \{ 3 \right \}$ is close to 1, suggesting that the event planes $\Psi _{2}$ and $\Psi _{4}$ gradually converge in the same direction at large $N$, consistent with the hydrodynamic expectation. The increase in $nsc_{2,4} \left \{ 4 \right \}$ and $nac_{2} \left \{ 3 \right \}$ with decreasing $N$ and the increase in $nsc_{2,4} \left \{ 4 \right \}$ and $nac_{2} \left \{ 3 \right \}$ with increasing $p$ are both due to the TMC effect.

\section{Conclusions}
In this paper, we calculated the four-particle symmetric cumulants $sc_{2,4} \left \{ 4 \right \}$, three-particle asymmetric cumulants $ac_{2} \left \{ 3 \right \}$, and the normalized cumulants $nsc_{2,4} \left \{ 4 \right \}$ and $nac_{2} \left \{ 3 \right \}$, originating from the transverse momentum conservation and flow. As expected, when the number of particles is small, the correlation comes from the TMC, and when the number of particles is large, the collective flow is dominant. Our results are consistent with the ATLAS data using the subevent cumulant method and therefore allow for a better understanding of collectivity in small systems. In the future, we can calculate the higher order symmetric cumulants $sc_{k,l,m}   \left \{ 6 \right \}$ in the same way to understand how the TMC and collective flow affects the coupling between $v_{k}$, $v_{l}$, and $v_{m}$ in small systems.

\section*{ACKNOWLEDGMENTS}
J.-L.P. thanks Mu-Ting Xie for helpful discussions. This work is partially supported by the National Natural Science Foundation of China under Grants  No. 12325507, No.12147101, No. 11890714, No. 11835002, No. 11961131011, No. 11421505, and No. 12105054, the National Key Research and Development Program of China under Grant No. 2022YFA1604900, the Strategic Priority Research Program of Chinese Academy of Sciences under Grant No. XDB34030000, and the Guangdong Major Project of Basic and Applied Basic Research under Grant No. 2020B0301030008 (J.-L.P. and G.-L.M.); and by the Ministry of Science and Higher Education (PL) and the National Science Centre (PL) under Grant No. 2018/30/Q/ST2/00101 (A.B.).

\setcounter{section}{0} 
\section*{APPENDIX}
\renewcommand{\thesection}{A} 
\setcounter{equation}{0} 
\renewcommand{\theequation}{\thesection.\arabic{equation}} 

The full results of the Eq.~\eqref{eq24} are as follows:
\begin{align*}
\label{eqA1}
A_{0} = \hspace{0.1cm}& v_{2}^{2} v_{4}^{2}, \notag \\
A_{1}=\hspace{0.1cm}&v_{2}^{2} v_{3}^{2} + 4 v_{2}^{2} v_{4}^{2} + v_{3}^{2} v_{4}^{2} - v_{2} v_{2F} v_{4}^{2} \cos(2 \Psi_{2}) + 2 v_{2} v_{3}^{2} v_{4} \cos(2 \Psi_{2} - 6 \Psi_{3} + 4 \Psi_{4})  
 - v_2^3 v_{2F} v_4 \cos(2 \Psi_2 - 4 \Psi_4) \notag \\& - 2 v_2 v_{2F} v_3^2 v_4 \cos(2 \Psi_2 - 4 \Psi_4) 
- v_2 v_{2F} v_4^3 \cos(2 \Psi_2 - 4 \Psi_4) ,\notag\\
A_{2} =\hspace{0.1cm}&v_{2}^{4} + 16 v_{2}^{2} v_{3}^{2} + 4 v_{3}^{4} + v_{4}^{2} + 30 v_{2}^{2} v_{4}^{2} + \frac{{v_{2F}^{2} v_{4}^{2}}}{2} + 16 v_{3}^{2} v_{4}^{2} + v_{4}^{4} - 6 v_{2} v_{2F} v_{3}^{2} \cos(2 \Psi_{2}) - 20 v_{2} v_{2F} v_{4}^{2} \cos(2 \Psi_{2}) \notag \\ &+ 2 v_{2}^{2} v_{4} \cos(4 \Psi_{2} - 4 \Psi_{4}) - 6 v_{2F} v_{3}^{2} v_{4} \cos(6 \Psi_{3} - 4 \Psi_{4}) + 24 v_{2} v_{3}^{2} v_{4} \cos(2 \Psi_{2} - 6 \Psi_{3} + 4 \Psi_{4})
 + \frac{v_2^4 v_{2F}^2}{2}  \notag\\&+ 8 v_2^2 v_{2F}^2 v_3^2  + 
2 v_{2F}^2 v_3^4  + \frac{v_{2F}^2 v_4^2}{2} + 
15 v_2^2 v_{2F}^2 v_4^2  + 8 v_{2F}^2 v_3^2 v_4^2  + \frac{v_{2F}^2 v_4^4}{2} + 
\frac{1}{2} v_2^2 v_{2F}^2 v_4^2 \cos(4 \Psi_2)\notag\\
&- 6 v_2^2 v_{2F} v_3^2 \cos(4 \Psi_2 - 6 \Psi_3) + 
3 v_2^2 v_{2F}^2 v_4^2 \cos(4 \Psi_2 - 8 \Psi_4) - 
6 v_{2F} v_3^2 v_4^2 \cos(6 \Psi_3 - 8 \Psi_4) \notag\\ &- 
20 v_2^3 v_{2F} v_4 \cos(2 \Psi_2 - 4 \Psi_4) - 
48 v_2 v_{2F} v_3^2 v_4 \cos(2 \Psi_2 - 4 \Psi_4) - 
20 v_2 v_{2F} v_4^3 \cos(2 \Psi_2 - 4 \Psi_4) \notag\\ & + 
v_2^2 v_{2F}^2 v_4 \cos(4 \Psi_2 - 4 \Psi_4)  + 
\frac{7}{2} v_2^2 v_{2F}^2 v_4 \cos(4 \Psi_4) + 
6 v_{2F}^2 v_3^2 v_4 \cos(4 \Psi_4) + 
3 v_{2F}^2 v_4^3 \cos(4 \Psi_4)  \notag\\&+ 
12 v_2 v_{2F}^2 v_3^2 v_4 \cos(2 \Psi_2 - 6 \Psi_3 + 4 \Psi_4),\notag\\
A_{3} =\hspace{0.1cm}&24 v_{2}^{4} + 9 v_{3}^{2} + 234 v_{2}^{2} v_{3}^{2} + \frac{{27 v_{2F}^{2} v_{3}^{2}}}{2} + 72 v_{3}^{4} + 24 v_{4}^{2} + 304 v_{2}^{2} v_{4}^{2} + 36 v_{2F}^{2} v_{4}^{2} + 234 v_{3}^{2} v_{4}^{2} + 24 v_{4}^{4} - 21 v_{2}^{3} v_{2F} \cos(2 \Psi_{2}) \notag \\ &- 210 v_{2} v_{2F} v_{3}^{2} \cos(2 \Psi_{2}) - 369 v_{2} v_{2F} v_{4}^{2} \cos(2 \Psi_{2}) - 21 v_{2} v_{2F} v_{4} \cos(2 \Psi_{2} - 4 \Psi_{4}) + 48 v_{2}^{2} v_{4} \cos(4 \Psi_{2} - 4 \Psi_{4}) \notag \\ &- 168 v_{2F} v_{3}^{2} v_{4} \cos(6 \Psi_{3} - 4 \Psi_{4}) + 300 v_{2} v_{3}^{2} v_{4} \cos(2 \Psi_{2} - 6 \Psi_{3} + 4 \Psi_{4})
 + 36 v_2^4 v_{2F}^2 + 351 v_2^2 v_{2F}^2 v_3^2  + 108 v_{2F}^2 v_3^4 \notag\\&+ 36 v_{2F}^2 v_4^2 + 456 v_2^2 v_{2F}^2 v_4^2 + 351 v_{2F}^2 v_3^2 v_4^2  + 36 v_{2F}^2 v_4^4 - 
\frac{21}{4} v_2^3 v_{2F}^3 \cos(2 \Psi_2)  - 
\frac{105}{2} v_2 v_{2F}^3 v_3^2 \cos(2 \Psi_2)  \notag\\&- 
\frac{369}{4} v_2 v_{2F}^3 v_4^2 \cos(2 \Psi_2)+ 
\frac{15}{2} v_2^2 v_{2F}^2 v_3^2 \cos(4 \Psi_2) + 
24 v_2^2 v_{2F}^2 v_4^2 \cos(4 \Psi_2) + 
\frac{105}{2} v_2 v_{2F}^2 v_3^2 \cos(2 \Psi_2 - 6 \Psi_3)\notag\\ &- 
168 v_2^2 v_{2F} v_3^2 \cos(4 \Psi_2 - 6 \Psi_3) - 
42 v_2^2 v_{2F}^3 v_3^2 \cos(4 \Psi_2 - 6 \Psi_3) - 
\frac{3}{2} v_{2F}^3 v_3^2 v_4^2 \cos(6 \Psi_3) - 
\frac{105}{4} v_2 v_{2F}^3 v_4^2 \cos(2 \Psi_2 - 8 \Psi_4)\notag\\ &+ 
120 v_2^2 v_{2F}^2 v_4^2 \cos(4 \Psi_2 - 8 \Psi_4) - 
168 v_{2F} v_3^2 v_4^2 \cos(6 \Psi_3 - 8 \Psi_4) - 
42 v_{2F}^3 v_3^2 v_4^2 \cos(6 \Psi_3 - 8 \Psi_4)\notag\\ & - 
363 v_2^3 v_{2F} v_4 \cos(2 \Psi_2 - 4 \Psi_4) - 
\frac{21}{4} v_2 v_{2F}^3 v_4 \cos(2 \Psi_2 - 4 \Psi_4) - 
\frac{363}{4} v_2^3 v_{2F}^3 v_4 \cos(2 \Psi_2 - 4 \Psi_4)\notag\\ & - 
936 v_2 v_{2F} v_3^2 v_4 \cos(2 \Psi_2 - 4 \Psi_4) - 
234 v_2 v_{2F}^3 v_3^2 v_4 \cos(2 \Psi_2 - 4 \Psi_4) - 
363 v_2 v_{2F} v_4^3 \cos(2 \Psi_2 - 4 \Psi_4)\notag\\ & - 
\frac{363}{4} v_2 v_{2F}^3 v_4^3 \cos(2 \Psi_2 - 4 \Psi_4)  + 
72 v_2^2 v_{2F}^2 v_4 \cos(4 \Psi_2 - 4 \Psi_4) - 
3 v_2^3 v_{2F} v_4 \cos(6 \Psi_2 - 4 \Psi_4) \notag\\ &- 
\frac{3}{4} v_2^3 v_{2F}^3 v_4 \cos(6 \Psi_2 - 4 \Psi_4)  - 
42 v_{2F}^3 v_3^2 v_4 \cos(6 \Psi_3 - 4 \Psi_4) + 
9 v_2 v_{2F}^2 v_3^2 v_4 \cos(2 \Psi_2 + 6 \Psi_3 - 4 \Psi_4) \notag\\ &+ 
144 v_2^2 v_{2F}^2 v_4 \cos(4 \Psi_4) + 
\frac{525}{2} v_{2F}^2 v_3^2 v_4 \cos(4 \Psi_4) + 
120 v_{2F}^2 v_4^3 \cos(4 \Psi_4) - 
\frac{15}{4} v_2 v_{2F}^3 v_4 \cos(2 \Psi_2 + 4 \Psi_4)\notag\\ & - 
4 v_2^3 v_{2F}^3 v_4 \cos(2 \Psi_2 + 4 \Psi_4) - 
9 v_2 v_{2F}^3 v_3^2 v_4 \cos(2 \Psi_2 + 4 \Psi_4) - 
4 v_2 v_{2F}^3 v_4^3 \cos(2 \Psi_2 + 4 \Psi_4) \notag\\ & + 
450 v_2 v_{2F}^2 v_3^2 v_4 \cos(2 \Psi_2 - 6 \Psi_3 + 4 \Psi_4),\notag\\
A_{4} =\hspace{0.1cm}&49 v_{2}^{2} + 436 v_{2}^{4} + 147 v_{2}^{2} v_{2F}^{2} + 264 v_{3}^{2} + 3328 v_{2}^{2} v_{3}^{2} + 792 v_{2F}^{2} v_{3}^{2} + 1056 v_{3}^{4} + 448 v_{4}^{2} + 3625 v_{2}^{2} v_{4}^{2} + 1344 v_{2F}^{2} v_{4}^{2} \notag \\ &+ 3328 v_{3}^{2} v_{4}^{2} + 436 v_{4}^{4} - 872 v_{2}^{3} v_{2F} \cos(2 \Psi_{2}) - 5004 v_{2} v_{2F} v_{3}^{2} \cos(2 \Psi_{2}) - 6728 v_{2} v_{2F} v_{4}^{2} \cos(2 \Psi_{2}) \notag \\ &+ 45 v_{2}^{2} v_{2F}^{2} \cos(4 \Psi_{2}) - 872 v_{2} v_{2F} v_{4} \cos(2 \Psi_{2} - 4 \Psi_{4}) + 898 v_{2}^{2} v_{4} \cos(4 \Psi_{2} - 4 \Psi_{4}) - 3492 v_{2F} v_{3}^{2} v_{4} \cos(6 \Psi_{3} - 4 \Psi_{4}) \notag \\ &+ 45 v_{2F}^{2} v_{4} \cos(4 \Psi_{4}) + 3920 v_{2} v_{3}^{2} v_{4} \cos(2 \Psi_{2} - 6 \Psi_{3} + 4 \Psi_{4})+ 1308 v_2^4 v_{2F}^2 + \frac{147 v_2^2 v_{2F}^4}{8} + \frac{327 v_2^4 v_{2F}^4}{2} + 9984 v_2^2 v_{2F}^2 v_3^2
\notag\\ &+ 99 v_{2F}^4 v_3^2 + 1248 v_2^2 v_{2F}^4 v_3^2  + 3168 v_{2F}^2 v_3^4 + 
396 v_{2F}^4 v_3^4  + 10875 v_2^2 v_{2F}^2 v_4^2 + 168 v_{2F}^4 v_4^2 + \frac{10875}{8} v_2^2 v_{2F}^4 v_4^2\notag\\ &+ 9984 v_{2F}^2 v_3^2 v_4^2 + 1248 v_{2F}^4 v_3^2 v_4^2  + 1308 v_{2F}^2 v_4^4 + \frac{327 v_{2F}^4 v_4^4}{2}- 
654 v_2^3 v_{2F}^3 \cos(2 \Psi_2)  - 
3753 v_2 v_{2F}^3 v_3^2 \cos(2 \Psi_2)\notag\\ &- 
5046 v_2 v_{2F}^3 v_4^2 \cos(2 \Psi_2) + 
48 v_2^4 v_{2F}^2 \cos(4 \Psi_2) + 
\frac{15}{2} v_2^2 v_{2F}^4 \cos(4 \Psi_2) + 
8 v_2^4 v_{2F}^4 \cos(4 \Psi_2)\notag\\ &+ 
510 v_2^2 v_{2F}^2 v_3^2 \cos(4 \Psi_2) + 
85 v_2^2 v_{2F}^4 v_3^2 \cos(4 \Psi_2) + 
837 v_2^2 v_{2F}^2 v_4^2 \cos(4 \Psi_2) + 
\frac{279}{2} v_2^2 v_{2F}^4 v_4^2 \cos(4 \Psi_2)\notag\\ &+ 
2196 v_2 v_{2F}^2 v_3^2 \cos(2 \Psi_2 - 6 \Psi_3) + 
366 v_2 v_{2F}^4 v_3^2 \cos(2 \Psi_2 - 6 \Psi_3) - 
3420 v_2^2 v_{2F} v_3^2 \cos(4 \Psi_2 - 6 \Psi_3) \notag\\ &- 
2565 v_2^2 v_{2F}^3 v_3^2 \cos(4 \Psi_2 - 6 \Psi_3) - 
105 v_{2F}^3 v_3^2 \cos(6 \Psi_3) - 
134 v_2^2 v_{2F}^3 v_3^2 \cos(6 \Psi_3) - 
42 v_{2F}^3 v_3^4 \cos(6 \Psi_3) 
\end{align*}

\begin{align*}
\hspace{0.1cm}&  - 
140 v_{2F}^3 v_3^2 v_4^2 \cos(6 \Psi_3) - 
1526 v_2 v_{2F}^3 v_4^2 \cos(2 \Psi_2 - 8 \Psi_4) + 
3270 v_2^2 v_{2F}^2 v_4^2 \cos(4 \Psi_2 - 8 \Psi_4) \notag\\ &+ 
545 v_2^2 v_{2F}^4 v_4^2 \cos(4 \Psi_2 - 8 \Psi_4) - 
3420 v_{2F} v_3^2 v_4^2 \cos(6 \Psi_3 - 8 \Psi_4) - 
2565 v_{2F}^3 v_3^2 v_4^2 \cos(6 \Psi_3 - 8 \Psi_4)  \notag\\ &- 
6424 v_2^3 v_{2F} v_4 \cos(2 \Psi_2 - 4 \Psi_4) - 
654 v_2 v_{2F}^3 v_4 \cos(2 \Psi_2 - 4 \Psi_4) - 
4818 v_2^3 v_{2F}^3 v_4 \cos(2 \Psi_2 - 4 \Psi_4) \notag\\ &- 
16640 v_2 v_{2F} v_3^2 v_4 \cos(2 \Psi_2 - 4 \Psi_4) - 
12480 v_2 v_{2F}^3 v_3^2 v_4 \cos(2 \Psi_2 - 4 \Psi_4) - 
6424 v_2 v_{2F} v_4^3 \cos(2 \Psi_2 - 4 \Psi_4) \notag\\ &- 
4818 v_2 v_{2F}^3 v_4^3 \cos(2 \Psi_2 - 4 \Psi_4)  + 
2694 v_2^2 v_{2F}^2 v_4 \cos(4 \Psi_2 - 4 \Psi_4) + 
\frac{1347}{4} v_2^2 v_{2F}^4 v_4 \cos(4 \Psi_2 - 4 \Psi_4) \notag\\ &- 
152 v_2^3 v_{2F} v_4 \cos(6 \Psi_2 - 4 \Psi_4) - 
114 v_2^3 v_{2F}^3 v_4 \cos(6 \Psi_2 - 4 \Psi_4) + 
22 v_2 v_{2F}^4 v_3^2 v_4 \cos(2 \Psi_2 - 6 \Psi_3 - 4 \Psi_4)  \notag\\ &- 
2619 v_{2F}^3 v_3^2 v_4 \cos(6 \Psi_3 - 4 \Psi_4) + 
510 v_2 v_{2F}^2 v_3^2 v_4 \cos(2 \Psi_2 + 6 \Psi_3 - 4 \Psi_4) + 
85 v_2 v_{2F}^4 v_3^2 v_4 \cos(2 \Psi_2 + 6 \Psi_3 - 4 \Psi_4)  \notag\\ &+ 
4419 v_2^2 v_{2F}^2 v_4 \cos(4 \Psi_4) + 
\frac{15}{2} v_{2F}^4 v_4 \cos(4 \Psi_4) + 
\frac{1473}{2} v_2^2 v_{2F}^4 v_4 \cos(4 \Psi_4) + 
7434 v_{2F}^2 v_3^2 v_4 \cos(4 \Psi_4) \notag\\ &+ 
1239 v_{2F}^4 v_3^2 v_4 \cos(4 \Psi_4) + 
3318 v_{2F}^2 v_4^3 \cos(4 \Psi_4) + 
553 v_{2F}^4 v_4^3 \cos(4 \Psi_4) + 
\frac{105}{2} v_{2F}^4 v_4^2 \cos(8 \Psi_4) + 
\frac{477}{8} v_2^2 v_{2F}^4 v_4^2 \cos(8 \Psi_4) \notag\\ &+ 
28 v_{2F}^4 v_3^2 v_4^2 \cos(8 \Psi_4) + 
\frac{7}{2} v_{2F}^4 v_4^4 \cos(8 \Psi_4) - 
266 v_2 v_{2F}^3 v_4 \cos(2 \Psi_2 + 4 \Psi_4) - 
288 v_2^3 v_{2F}^3 v_4 \cos(2 \Psi_2 + 4 \Psi_4) \notag\\ &- 
694 v_2 v_{2F}^3 v_3^2 v_4 \cos(2 \Psi_2 + 4 \Psi_4) - 
288 v_2 v_{2F}^3 v_4^3 \cos(2 \Psi_2 + 4 \Psi_4) + 
\frac{49}{4} v_2^2 v_{2F}^4 v_4 \cos(4 \Psi_2 + 4 \Psi_4) \notag\\ & + 
11760 v_2 v_{2F}^2 v_3^2 v_4 \cos(2 \Psi_2 - 6 \Psi_3 + 4 \Psi_4) + 
1470 v_2 v_{2F}^4 v_3^2 v_4 \cos(2 \Psi_2 - 6 \Psi_3 + 4 \Psi_4)
,\notag\\
A_{5} =\hspace{0.1cm}&1820 v_{2}^{2} + 7120 v_{2}^{4} + 9100 v_{2}^{2} v_{2F}^{2} + 5545 v_{3}^{2} + 47025 v_{2}^{2} v_{3}^{2} + 27725 v_{2F}^{2} v_{3}^{2} + 14800 v_{3}^{4} + 7680 v_{4}^{2} + 47004 v_{2}^{2} v_{4}^{2} \notag \\ &+ 38400 v_{2F}^{2} v_{4}^{2} + 47001 v_{3}^{2} v_{4}^{2} + 7120 v_{4}^{4} - 525 v_{2} v_{2F} \cos(2 \Psi_{2}) - 23855 v_{2}^{3} v_{2F} \cos(2 \Psi_{2}) - \frac{1575}{2} v_{2} v_{2F}^{3} \cos(2 \Psi_{2}) \notag \\ &- 102960 v_{2} v_{2F} v_{3}^{2} \cos(2 \Psi_{2}) - 120970 v_{2} v_{2F} v_{4}^{2} \cos(2 \Psi_{2}) + 2940 v_{2}^{2} v_{2F}^{2} \cos(4 \Psi_{2}) - 23895 v_{2} v_{2F} v_{4} \cos(2 \Psi_{2} - 4 \Psi_{4}) \notag \\ &+ 15480 v_{2}^{2} v_{4} \cos(4 \Psi_{2} - 4 \Psi_{4}) - 65430 v_{2F} v_{3}^{2} v_{4} \cos(6 \Psi_{3} - 4 \Psi_{4}) + 2940 v_{2F}^{2} v_{4} \cos(4 \Psi_{4}) \notag \\ &+ 52730 v_{2} v_{3}^{2} v_{4} \cos(2 \Psi_{2} - 6 \Psi_{3} + 4 \Psi_{4})
+ 35600 v_{2}^{4} v_{2F}^{2} + \frac{6825 v_{2}^{2} v_{2F}^{4}}{2} + 13350 v_{2}^{4} v_{2F}^{4} + 235125 v_{2}^{2} v_{2F}^{2} v_{3}^{2} + \frac{83175 v_{2F}^{4} v_{3}^{2}}{8} \notag\\ &+ \frac{705375}{8} v_{2}^{2} v_{2F}^{4} v_{3}^{2}  + 74000 v_{2F}^{2} v_{3}^{4} + 27750 v_{2F}^{4} v_{3}^{4}   + 235020 v_{2}^{2} v_{2F}^{2} v_{4}^{2} + 14400 v_{2F}^{4} v_{4}^{2} + \frac{176265 v_{2}^{2}}{2} v_{2F}^{4} v_{4}^{2}  \notag\\ &+ 
235005 v_{2F}^{2} v_{3}^{2} v_{4}^{2} + \frac{705015}{8} v_{2F}^{4} v_{3}^{2} v_{4}^{2}+ 35600 v_{2F}^{2} v_{4}^{4} + 13350 v_{2F}^{4} v_{4}^{4}- 
\frac{71565}{2}  v_{2}^{3} v_{2F}^{3} \cos(2 \Psi_{2})- \frac{525}{8} v_{2} v_{2F}^{5} \cos(2 \Psi_{2})\notag\\ & - \frac{23855}{8} v_{2}^{3} v_{2F}^{5} \cos(2 \Psi_{2})- 154440 v_{2} v_{2F}^{3} v_{3}^{2} \cos(2 \Psi_{2}) - 12870 v_{2} v_{2F}^{5} v_{3}^{2} \cos(2 \Psi_{2})  - 181455 v_{2} v_{2F}^{3} v_{4}^{2} \cos(2 \Psi_{2})\notag\\ &- \frac{60485}{4} v_{2} v_{2F}^{5} v_{4}^{2} \cos(2 \Psi_{2})  + 
3200 v_{2}^{4} v_{2F}^{2} \cos(4 \Psi_{2}) + 1470 v_{2}^{2} v_{2F}^{4} \cos(4 \Psi_{2}) + 1600 v_{2}^{4} v_{2F}^{4} \cos(4 \Psi_{2})\notag\\ &+ 19785 v_{2}^{2} v_{2F}^{2} v_{3}^{2} \cos(4 \Psi_{2}) + \frac{19785}{2} v_{2}^{2} v_{2F}^{4} v_{3}^{2} \cos(4 \Psi_{2}) + 
24780 v_{2}^{2} v_{2F}^{2} v_{4}^{2} \cos(4 \Psi_{2}) + 12390 v_{2}^{2} v_{2F}^{4} v_{4}^{2} \cos(4 \Psi_{2}) \notag\\ &- 245 v_{2}^{3} v_{2F}^{3} \cos(6 \Psi_{2}) - \frac{245}{8} v_{2}^{3} v_{2F}^{5} \cos(6 \Psi_{2}) + 61755 v_{2} v_{2F}^{2} v_{3}^{2} \cos(2 \Psi_{2} - 6 \Psi_{3}) + 
\frac{61755}{2} v_{2} v_{2F}^{4} v_{3}^{2} \cos(2 \Psi_{2} - 6 \Psi_{3}) \notag\\ &- 61920 v_{2}^{2} v_{2F} v_{3}^{2} \cos(4 \Psi_{2} - 6 \Psi_{3}) - 92880 v_{2}^{2} v_{2F}^{3} v_{3}^{2} \cos(4 \Psi_{2} - 6 \Psi_{3}) - \
7740 v_{2}^{2} v_{2F}^{5} v_{3}^{2} \cos(4 \Psi_{2} - 6 \Psi_{3}) \notag\\ &- 6440 v_{2F}^{3} v_{3}^{2} \cos(6 \Psi_{3}) - 9265 v_{2}^{2} v_{2F}^{3} v_{3}^{2} \cos(6 \Psi_{3}) - 805 v_{2F}^{5} v_{3}^{2} \cos(6 \Psi_{3}) - \
\frac{9265}{8} v_{2}^{2} v_{2F}^{5} v_{3}^{2} \cos(6 \Psi_{3})\notag\\ &- 2840 v_{2F}^{3} v_{3}^{4} \cos(6 \Psi_{3})- 355 v_{2F}^{5} v_{3}^{4} \cos(6 \Psi_{3}) - 7360 v_{2F}^{3} v_{3}^{2} v_{4}^{2} \cos(6 \Psi_{3}) - \
920 v_{2F}^{5} v_{3}^{2} v_{4}^{2} \cos(6 \Psi_{3})\notag\\ &+ \frac{1305}{2} v_{2} v_{2F}^{4} v_{3}^{2} \cos(2 \Psi_{2} + 6 \Psi_{3})- \frac{1305}{16} v_{2} v_{2F}^{5} v_{4}^{3} \cos(2 \Psi_{2} - 12 \Psi_{4}) - 
\frac{108855}{2} v_{2} v_{2F}^{3} v_{4}^{2} \cos(2 \Psi_{2} - 8 \Psi_{4})\notag\\ &- \frac{108855}{16} v_{2} v_{2F}^{5} v_{4}^{2} \cos(2 \Psi_{2} - 8 \Psi_{4})+ 74760 v_{2}^{2} v_{2F}^{2} v_{4}^{2} \cos(4 \Psi_{2} - 8 \Psi_{4}) + \
37380 v_{2}^{2} v_{2F}^{4} v_{4}^{2} \cos(4 \Psi_{2} - 8 \Psi_{4})\notag\\ &- 61920 v_{2F} v_{3}^{2} v_{4}^{2} \cos(6 \Psi_{3} - 8 \Psi_{4})- 92880 v_{2F}^{3} v_{3}^{2} v_{4}^{2} \cos(6 \Psi_{3} - 8 \Psi_{4}) - \
7740 v_{2F}^{5} v_{3}^{2} v_{4}^{2} \cos(6 \Psi_{3} - 8 \Psi_{4})\notag\\ &- 110770 v_{2}^{3} v_{2F} v_{4} \cos(2 \Psi_{2} - 4 \Psi_{4}) - \
\frac{71685}{2} v_{2} v_{2F}^{3} v_{4} \cos(2 \Psi_{2} - 4 \Psi_{4}) - 166155 v_{2}^{3} v_{2F}^{3} v_{4} \cos(2 \Psi_{2} - 4 \Psi_{4})\notag\\ &- \frac{23895}{8} v_{2} v_{2F}^{5} v_{4} \cos(2 \Psi_{2} - 4 \Psi_{4}) - 
\frac{55385}{4} v_{2}^{3} v_{2F}^{5} v_{4} \cos(2 \Psi_{2} - 4 \Psi_{4}) - 282030 v_{2} v_{2F} v_{3}^{2} v_{4} \cos(2 \Psi_{2} - 4 \Psi_{4})\notag\\ &- 423045 v_{2} v_{2F}^{3} v_{3}^{2} v_{4} \cos(2 \Psi_{2} - 4 \Psi_{4}) - \
\frac{141015}{4} v_{2} v_{2F}^{5} v_{3}^{2} v_{4} \cos(2 \Psi_{2} - 4 \Psi_{4}) - 110695 v_{2} v_{2F} v_{4}^{3} \cos(2 \Psi_{2} - 4 \Psi_{4})\notag\\ &- \frac{332085}{2} v_{2} v_{2F}^{3} v_{4}^{3} \cos(2 \Psi_{2} - 4 \Psi_{4}) - \
\frac{110695}{8} v_{2} v_{2F}^{5} v_{4}^{3} \cos(2 \Psi_{2} - 4 \Psi_{4})+ 77400 v_{2}^{2} v_{2F}^{2} v_{4} \cos(4 \Psi_{2} - 4 \Psi_{4})\notag\\ &+ 
29025 v_{2}^{2} v_{2F}^{4} v_{4} \cos(4 \Psi_{2} - 4 \Psi_{4}) - 4875 v_{2}^{3} v_{2F} v_{4} \cos(6 \Psi_{2} - 4 \Psi_{4})- \frac{14625}{2} v_{2}^{3} v_{2F}^{3} v_{4} \cos(6 \Psi_{2} - 4 \Psi_{4})
\end{align*}

\begin{align*}
  \hspace{1cm}&  - 
\frac{4875}{8} v_{2}^{3} v_{2F}^{5} v_{4} \cos(6 \Psi_{2} - 4 \Psi_{4}) + \frac{10665}{4} v_{2} v_{2F}^{4} v_{3}^{2} v_{4} \cos(2 \Psi_{2} - 6 \Psi_{3} - 4 \Psi_{4})  - 98145 v_{2F}^{3} v_{3}^{2} v_{4} \cos(6 \Psi_{3} - 4 \Psi_{4}) \notag\\ &- \frac{32715}{4} v_{2F}^{5} v_{3}^{2} v_{4} \cos(6 \Psi_{3} - 4 \Psi_{4})+ 18300 v_{2} v_{2F}^{2} v_{3}^{2} v_{4} \cos(2 \Psi_{2} + 6 \Psi_{3} - 4 \Psi_{4}) + \
9150 v_{2} v_{2F}^{4} v_{3}^{2} v_{4} \cos(2 \Psi_{2} + 6 \Psi_{3} - 4 \Psi_{4})  \notag\\ &+ 114820 v_{2}^{2} v_{2F}^{2} v_{4} \cos(4 \Psi_{4}) + \
1470 v_{2F}^{4} v_{4} \cos(4 \Psi_{4}) + 57410 v_{2}^{2} v_{2F}^{4} v_{4} \cos(4 \Psi_{4}) + 175995 v_{2F}^{2} v_{3}^{2} v_{4} \cos(4 \Psi_{4}) \notag\\ &+ \frac{175995}{2} v_{2F}^{4} v_{3}^{2} v_{4} \cos(4 \Psi_{4}) + \
77960 v_{2F}^{2} v_{4}^{3} \cos(4 \Psi_{4}) + 38980 v_{2F}^{4} v_{4}^{3} \cos(4 \Psi_{4}) + 4410 v_{2F}^{4} v_{4}^{2} \cos(8 \Psi_{4}) \notag\\ &+ \frac{10415}{2} v_{2}^{2} v_{2F}^{4} v_{4}^{2} \cos(8 \Psi_{4}) + \
\frac{22005}{8} v_{2F}^{4} v_{3}^{2} v_{4}^{2} \cos(8 \Psi_{4}) + 390 v_{2F}^{4} v_{4}^{4} \cos(8 \Psi_{4}) - 12600 v_{2} v_{2F}^{3} v_{4} \cos(2 \Psi_{2} + 4 \Psi_{4}) \notag\\ &- 
13670 v_{2}^{3} v_{2F}^{3} v_{4} \cos(2 \Psi_{2} + 4 \Psi_{4}) - 1575 v_{2} v_{2F}^{5} v_{4} \cos(2 \Psi_{2} + 4 \Psi_{4}) - \frac{6835}{4} v_{2}^{3} v_{2F}^{5} v_{4} \cos(2 \Psi_{2} + 4 \Psi_{4}) \notag\\ &- \
31010 v_{2} v_{2F}^{3} v_{3}^{2} v_{4} \cos(2 \Psi_{2} + 4 \Psi_{4}) - \frac{15505}{4} v_{2} v_{2F}^{5} v_{3}^{2} v_{4} \cos(2 \Psi_{2} + 4 \Psi_{4}) - 12620 v_{2} v_{2F}^{3} v_{4}^{3} \cos(2 \Psi_{2} + 4 \Psi_{4}) \notag\\ &- \
\frac{3155}{2} v_{2} v_{2F}^{5} v_{4}^{3} \cos(2 \Psi_{2} + 4 \Psi_{4}) + 1335 v_{2}^{2} v_{2F}^{4} v_{4} \cos(4 \Psi_{2} + 4 \Psi_{4}) - \frac{105}{8} v_{2}^{3} v_{2F}^{5} v_{4} \cos(6 \Psi_{2} + 4 \Psi_{4}) \notag\\ & + 263650 v_{2} v_{2F}^{2} v_{3}^{2} v_{4} \cos(2 \Psi_{2} - 6 \Psi_{3} + 4 \Psi_{4}) + \
\frac{395475}{4} v_{2} v_{2F}^{4} v_{3}^{2} v_{4} \cos(2 \Psi_{2} - 6 \Psi_{3} + 4 \Psi_{4}) \notag\\ &- \frac{405}{4} v_{2F}^{5} v_{3}^{2} v_{4} \cos(6 \Psi_{3} + 4 \Psi_{4}) - \frac{4665}{16} v_{2} v_{2F}^{5} v_{4}^{2} \cos(2 \Psi_{2} + 8 \Psi_{4}),\notag\\ 
A_{6} =\hspace{0.1cm}&225 + 45396 v_{2}^{2} + 110941 v_{2}^{4} + \frac{{3375 v_{2F}^{2}}}{2} + 340470 v_{2}^{2} v_{2F}^{2} + \frac{{10125 v_{2F}^{4}}}{8} + 102780 v_{3}^{2} + 666216 v_{2}^{2} v_{3}^{2} \notag \\ &+ 770850 v_{2F}^{2} v_{3}^{2} + 205956 v_{3}^{4} + 126720 v_{4}^{2}+637704v_{2}^{2}v_{4}^{2}+950400v_{2F}^{2}v_{4}^{2} + 664800 v_{3}^{2} v_{4}^{2} + 110686 v_{4}^{4} \notag \\ &- 28140 v_{2} v_{2F} \cos(2 \Psi_{2}) - 545892 v_{2}^{3} v_{2F} \cos(2 \Psi_{2}) - 70350 v_{2} v_{2F}^{3} \cos(2 \Psi_{2}) - 1970532 v_{2} v_{2F} v_{3}^{2} \cos(2 \Psi_{2}) \notag \\ &- 2142936 v_{2} v_{2F} v_{4}^{2} \cos(2 \Psi_{2}) + 115605 v_{2}^{2} v_{2F}^{2} \cos(4 \Psi_{2}) - 548580 v_{2} v_{2F} v_{4} \cos(2 \Psi_{2} - 4 \Psi_{4}) \notag \\ &+ 257412 v_{2}^{2} v_{4} \cos(4 \Psi_{2} - 4 \Psi_{4}) - 1169892 v_{2F} v_{3}^{2} v_{4} \cos(6 \Psi_{3} - 4 \Psi_{4}) + 113610 v_{2F}^{2} v_{4} \cos(4 \Psi_{4}) \notag \\ &+ 724092 v_{2} v_{3}^{2} v_{4} \cos(2 \Psi_{2} - 6 \Psi_{3} + 4 \Psi_{4})
  + \frac{1664115 v_2^4 v_{2F}^2}{2} + \frac{510705 v_2^2 v_{2F}^4}{2} + \frac{4992345 v_2^4 v_{2F}^4}{8}\notag\\ &+ \frac{1125 v_{2F}^6}{16} + \frac{56745 v_2^2 v_{2F}^6}{4} + \frac{554705 v_2^4 v_{2F}^6}{16} + 4996620 v_2^2 v_{2F}^2 v_3^2+ \frac{1156275 v_{2F}^4 v_3^2}{2} + 3747465 v_2^2 v_{2F}^4 v_3^2 \notag\\ &+ \frac{128475 v_{2F}^6 v_3^2}{4} + \frac{416385}{2} v_2^2 v_{2F}^6 v_3^2+ 1544670 v_{2F}^2 v_3^4+ \frac{2317005 v_{2F}^4 v_3^4}{2} + \frac{257445 v_{2F}^6 v_3^4}{4}   + 4782780 v_2^2 v_{2F}^2 v_4^2\notag\\ &+ 712800 v_{2F}^4 v_4^2 + 3587085 v_2^2 v_{2F}^4 v_4^2+ 39600 v_{2F}^6 v_4^2 + \frac{398565}{2} v_2^2 v_{2F}^6 v_4^2 + 4986000 v_{2F}^2 v_3^2 v_4^2 \notag\\ &+ 3739500 v_{2F}^4 v_3^2 v_4^2+ 207750 v_{2F}^6 v_3^2 v_4^2  + 830145 v_{2F}^2 v_4^4 + \frac{2490435 v_{2F}^4 v_4^4}{4}+ \frac{276715 v_{2F}^6 v_4^4}{8}- 1364730 v_2^3 v_{2F}^3 \cos(2 \Psi_2)\notag\\ &- \frac{35175}{2} v_2 v_{2F}^5 \cos(2 \Psi_2)- \frac{682365}{2} v_2^3 v_{2F}^5 \cos(2 \Psi_2) -4926330 v_2 v_{2F}^3 v_3^2 \cos(2 \Psi_2)- \frac{2463165}{2} v_2 v_{2F}^5 v_3^2 \cos(2 \Psi_2)\notag\\ &- 5357340 v_2 v_{2F}^3 v_4^2 \cos(2 \Psi_2)- 1339335 v_2 v_{2F}^5 v_4^2 \cos(2 \Psi_2)+ 123315 v_2^4 v_{2F}^2 \cos(4 \Psi_2) + 115605 v_2^2 v_{2F}^4 \cos(4 \Psi_2)\notag\\ & + 123315 v_2^4 v_{2F}^4 \cos(4 \Psi_2) + \frac{115605}{16} v_2^2 v_{2F}^6 \cos(4 \Psi_2) + \frac{123315}{16} v_2^4 v_{2F}^6 \cos(4 \Psi_2) + 592860 v_2^2 v_{2F}^2 v_3^2 \cos(4 \Psi_2)\notag\\ &+ 592860 v_2^2 v_{2F}^4 v_3^2 \cos(4 \Psi_2)+ \frac{148215}{4} v_2^2 v_{2F}^6 v_3^2 \cos(4 \Psi_2) + \frac{1297635}{2} v_2^2 v_{2F}^2 v_4^2 \cos(4 \Psi_2)+ \frac{1297635}{2} v_2^2 v_{2F}^4 v_4^2 \cos(4 \Psi_2)\notag\\ &+ \frac{1297635}{32} v_2^2 v_{2F}^6 v_4^2 \cos(4 \Psi_2)- 23220 v_2^3 v_{2F}^3 \cos(6 \Psi_2) - \frac{17415}{2} v_2^3 v_{2F}^5 \cos(6 \Psi_2)+ \frac{1575}{4} v_2^4 v_{2F}^4 \cos(8 \Psi_2)\notag\\ &+ \frac{315}{8} v_2^4 v_{2F}^6 \cos(8 \Psi_2) + 1467090 v_2 v_{2F}^2 v_3^2 \cos(2 \Psi_2 - 6 \Psi_3)+ 1467090 v_2 v_{2F}^4 v_3^2 \cos(2 \Psi_2 - 6 \Psi_3)\notag\\ &+\frac{733545}{8} v_2 v_{2F}^6 v_3^2 \cos(2 \Psi_2 - 6 \Psi_3)- 1061160 v_2^2 v_{2F} v_3^2 \cos(4 \Psi_2 - 6 \Psi_3) - 2652900 v_2^2 v_{2F}^3 v_3^2 \cos(4 \Psi_2 - 6 \Psi_3)\notag\\ &- 663225 v_2^2 v_{2F}^5 v_3^2 \cos(4 \Psi_2 - 6 \Psi_3)- 246240 v_{2F}^3 v_3^2 \cos(6 \Psi_3)- 388860 v_2^2 v_{2F}^3 v_3^2 \cos(6 \Psi_3)\notag\\ & - 92340 v_{2F}^5 v_3^2 \cos(6 \Psi_3)- \frac{291645}{2} v_2^2 v_{2F}^5 v_3^2 \cos(6 \Psi_3)- 117990 v_{2F}^3 v_3^4 \cos(6 \Psi_3) - \frac{176985}{4} v_{2F}^5 v_3^4 \cos(6 \Psi_3)\notag\\ &- 285210 v_{2F}^3 v_3^2 v_4^2 \cos(6 \Psi_3) - \frac{427815}{4} v_{2F}^5 v_3^2 v_4^2 \cos(6 \Psi_3)+ \frac{246105}{4} v_2 v_{2F}^4 v_3^2 \cos(2 \Psi_2 + 6 \Psi_3)\notag\\ &+ \frac{49221}{8} v_2 v_{2F}^6 v_3^2 \cos(2 \Psi_2 + 6 \Psi_3)- \frac{4653}{4} v_2^2 v_{2F}^5 v_3^2 \cos(4 \Psi_2 + 6 \Psi_3)- \frac{39435}{4} v_2 v_{2F}^5 v_4^3 \cos(2 \Psi_2 - 12 \Psi_4)\notag\\ &- 1552110 v_2 v_{2F}^3 v_4^2 \cos(2 \Psi_2 - 8 \Psi_4)- \frac{2328165}{4} v_2 v_{2F}^5 v_4^2 \cos(2 \Psi_2 - 8 \Psi_4)+ \frac{3099615}{2} v_2^2 v_{2F}^2 v_4^2 \cos(4 \Psi_2 - 8 \Psi_4)
\end{align*}

\begin{align}
 \hspace{1cm}&  
 + \frac{3099615}{2} v_2^2 v_{2F}^4 v_4^2 \cos(4 \Psi_2 - 8 \Psi_4) + \frac{3099615}{32} v_2^2 v_{2F}^6 v_4^2 \cos(4 \Psi_2 - 8 \Psi_4)- 1060542 v_{2F} v_3^2 v_4^2 \cos(6 \Psi_3 - 8 \Psi_4)\notag\\ &- 2651355 v_{2F}^3 v_3^2 v_4^2 \cos(6 \Psi_3 - 8 \Psi_4) - \frac{2651355}{4} v_{2F}^5 v_3^2 v_4^2 \cos(6 \Psi_3 - 8 \Psi_4)- 1870296 v_2^3 v_{2F} v_4 \cos(2 \Psi_2 - 4 \Psi_4) \notag\\ &- 1371450 v_2 v_{2F}^3 v_4 \cos(2 \Psi_2 - 4 \Psi_4)- 4675740 v_2^3 v_{2F}^3 v_4 \cos(2 \Psi_2 - 4 \Psi_4) - \frac{685725}{2} v_2 v_{2F}^5 v_4 \cos(2 \Psi_2 - 4 \Psi_4) \notag\\ &- 1168935 v_2^3 v_{2F}^5 v_4 \cos(2 \Psi_2 - 4 \Psi_4)- 4655652 v_2 v_{2F} v_3^2 v_4 \cos(2 \Psi_2 - 4 \Psi_4) - 11639130 v_2 v_{2F}^3 v_3^2 v_4 \cos(2 \Psi_2 - 4 \Psi_4)\notag\\ &- \frac{5819565}{2} v_2 v_{2F}^5 v_3^2 v_4 \cos(2 \Psi_2 - 4 \Psi_4)- 1864836 v_2 v_{2F} v_4^3 \cos(2 \Psi_2 - 4 \Psi_4) - 4662090 v_2 v_{2F}^3 v_4^3 \cos(2 \Psi_2 - 4 \Psi_4) \notag\\ &- \frac{2331045}{2} v_2 v_{2F}^5 v_4^3 \cos(2 \Psi_2 - 4 \Psi_4) + 1930590 v_2^2 v_{2F}^2 v_4 \cos(4 \Psi_2 - 4 \Psi_4) + \frac{2895885}{2} v_2^2 v_{2F}^4 v_4 \cos(4 \Psi_2 - 4 \Psi_4)\notag\\ &+ \frac{321765}{4} v_2^2 v_{2F}^6 v_4 \cos(4 \Psi_2 - 4 \Psi_4) - 124980 v_2^3 v_{2F} v_4 \cos(6 \Psi_2 - 4 \Psi_4) - 312450 v_2^3 v_{2F}^3 v_4 \cos(6 \Psi_2 - 4 \Psi_4)\notag\\ &- \frac{156225}{2} v_2^3 v_{2F}^5 v_4 \cos(6 \Psi_2 - 4 \Psi_4) + \frac{635625}{4} v_2 v_{2F}^4 v_3^2 v_4 \cos(2 \Psi_2 - 6 \Psi_3 - 4 \Psi_4) \notag\\ &+ \frac{127125}{8} v_2 v_{2F}^6 v_3^2 v_4 \cos(2 \Psi_2 - 6 \Psi_3 - 4 \Psi_4) 
- 2924730 v_{2F}^3 v_3^2 v_4 \cos(6 \Psi_3 - 4 \Psi_4)  - \frac{1462365}{2} v_{2F}^5 v_3^2 v_4 \cos(6 \Psi_3 - 4 \Psi_4) 
\notag\\ &+ 528240 v_2 v_{2F}^2 v_3^2 v_4 \cos(2 \Psi_2 + 6 \Psi_3 - 4 \Psi_4)  + 528240 v_2 v_{2F}^4 v_3^2 v_4 \cos(2 \Psi_2 + 6 \Psi_3 - 4 \Psi_4)  \notag\\ &+ 33015 v_2 v_{2F}^6 v_3^2 v_4 \cos(2 \Psi_2 + 6 \Psi_3 - 4 \Psi_4) 
+ 2678325 v_2^2 v_{2F}^2 v_4 \cos(4 \Psi_4) + 113610 v_{2F}^4 v_4 \cos(4 \Psi_4)  \notag\\ &+ 2678325 v_2^2 v_{2F}^4 v_4 \cos(4 \Psi_4) 
 + \frac{56805}{8} v_{2F}^6 v_4 \cos(4 \Psi_4) + \frac{2678325}{16} v_2^2 v_{2F}^6 v_4 \cos(4 \Psi_4) + 3789540 v_{2F}^2 v_3^2 v_4 \cos(4 \Psi_4) \notag\\ &+ 3789540 v_{2F}^4 v_3^2 v_4 \cos(4 \Psi_4) 
 + \frac{947385}{4} v_{2F}^6 v_3^2 v_4 \cos(4 \Psi_4) + 1673160 v_{2F}^2 v_4^3 \cos(4 \Psi_4) + 1673160 v_{2F}^4 v_4^3 \cos(4 \Psi_4) \notag\\ &+ \frac{209145}{2} v_{2F}^6 v_4^3 \cos(4 \Psi_4) 
 + \frac{1660725}{8} v_{2F}^4 v_4^2 \cos(8 \Psi_4) + \frac{2073165}{8} v_2^2 v_{2F}^4 v_4^2 \cos(8 \Psi_4) + \frac{332145}{16} v_{2F}^6 v_4^2 \cos(8 \Psi_4) \notag\\ &+ \frac{414633}{16} v_2^2 v_{2F}^6 v_4^2 \cos(8 \Psi_4) 
 + \frac{602025}{4} v_{2F}^4 v_3^2 v_4^2 \cos(8 \Psi_4) + \frac{120405}{8} v_{2F}^6 v_3^2 v_4^2 \cos(8 \Psi_4) + \frac{48375}{2} v_{2F}^4 v_4^4 \cos(8 \Psi_4)\notag\\ &+\frac{9675}{4} v_{2F}^6 v_4^4 \cos(8 \Psi_4) + \frac{8415}{16} v_{2F}^6 v_4^3 \cos(12 \Psi_4) - 459360 v_2 v_{2F}^3 v_4 \cos(2 \Psi_2 + 4 \Psi_4)\notag\\&- 497720 v_2^3 v_{2F}^3 v_4 \cos(2 \Psi_2 + 4 \Psi_4)  
 - 172260 v_2 v_{2F}^5 v_4 \cos(2 \Psi_2 + 4 \Psi_4) - 186645 v_2^3 v_{2F}^5 v_4 \cos(2 \Psi_2 + 4 \Psi_4) 
 \notag\\& - 1064970 v_2 v_{2F}^3 v_3^2 v_4 \cos(2 \Psi_2 + 4 \Psi_4) - \frac{1597455}{4} v_2 v_{2F}^5 v_3^2 v_4 \cos(2 \Psi_2 + 4 \Psi_4) - 429680 v_2 v_{2F}^3 v_4^3 \cos(2 \Psi_2 + 4 \Psi_4) 
 \notag\\ &- 161130 v_2 v_{2F}^5 v_4^3 \cos(2 \Psi_2 + 4 \Psi_4) + \frac{357255}{4} v_2^2 v_{2F}^4 v_4 \cos(4 \Psi_2 + 4 \Psi_4) + \frac{71451}{8} v_2^2 v_{2F}^6 v_4 \cos(4 \Psi_2 + 4 \Psi_4) 
 \notag\\ &- \frac{4059}{2} v_2^3 v_{2F}^5 v_4 \cos(6 \Psi_2 + 4 \Psi_4) + 5430690 v_2 v_{2F}^2 v_3^2 v_4 \cos(2 \Psi_2 - 6 \Psi_3 + 4 \Psi_4) 
\notag\\ & + \frac{8146035}{2} v_2 v_{2F}^4 v_3^2 v_4 \cos(2 \Psi_2 - 6 \Psi_3 + 4 \Psi_4) + \frac{905115}{4} v_2 v_{2F}^6 v_3^2 v_4 \cos(2 \Psi_2 - 6 \Psi_3 + 4 \Psi_4) \notag\\ &- \frac{68013}{4} v_{2F}^5 v_3^2 v_4 \cos(6 \Psi_3 + 4 \Psi_4) 
 + \frac{1287}{4} v_2 v_{2F}^6 v_3^2 v_4 \cos(2 \Psi_2 + 6 \Psi_3 + 4 \Psi_4)\notag\\ & - \frac{137643}{4} v_2 v_{2F}^5 v_4^2 \cos(2 \Psi_2 + 8 \Psi_4) + \frac{15279}{32} v_2^2 v_{2F}^6 v_4^2 \cos(4 \Psi_2 + 8 \Psi_4) .
\end{align}

The full results of the Eq.~\eqref{eq32} are as follows:
\begin{align}
\label{eqA2}
D_{0} = \hspace{0.1cm}&v_2^2 v_4 \cos(4 \Psi_2 - 4 \Psi_4),\notag\\
D_{1} = \hspace{0.1cm}&- v_{2}v_{2F}v_{4}\cos(2\Psi_{2}-4\Psi_{4})
+3 v_{2}^{2}v_{4}\cos(4\Psi_{2} -4\Psi_{4} )-\frac{1}{2} v_2^3 v_{2F} \cos(2 \Psi_2) - 2 v_2 v_{2F} v_3^2 \cos(2 \Psi_2) 
\notag\\&- v_2 v_{2F} v_4^2 \cos(2 \Psi_2) \- v_{2F} v_3^2 v_4 \cos(6 \Psi_3 - 4 \Psi_4)
,\notag\\ 
D_{2} = \hspace{0.1cm}&2 v_{2}^{2}+4v_{3}^{2}+2 v_{4}^{2}-14 v_2 v_{2F} v_4 \cos(2 \Psi_2 - 4 \Psi_4)+15v_{2}^{2}v_{4}\cos(4\Psi_{2} -4\Psi_{4} )+\frac{3}{2}v_{2F}^{2}v_{4}\cos(4\Psi _{4} )   + v_2^2 v_{2F}^2  + 2 v_{2F}^2 v_3^2 \notag\\& + v_{2F}^2 v_4^2 - 7 v_2^3 v_{2F} \cos(2 \Psi_2) - 24 v_2 v_{2F} v_3^2 \cos(2 \Psi_2) - 14 v_2 v_{2F} v_4^2 \cos(2 \Psi_2) + \frac{1}{4} v_2^2 v_{2F}^2 \cos(4 \Psi_2)\notag\\ & + 3 v_2 v_{2F}^2 v_3^2 \cos(2 \Psi_2 - 6 \Psi_3) + \frac{15}{2} v_2^2 v_{2F}^2 v_4 \cos(4 \Psi_2 - 4 \Psi_4) - 12 v_{2F} v_3^2 v_4 \cos(6 \Psi_3 - 4 \Psi_4)  \notag\\ &+ \frac{7}{2} v_2^2 v_{2F}^2 v_4 \cos(4 \Psi_4) + 6 v_{2F}^2 v_3^2 v_4 \cos(4 \Psi_4) + \frac{3}{2} v_{2F}^2 v_4^3 \cos(4 \Psi_4),\notag\\ 
D_{3} = \hspace{0.1cm}&30v_{2}^{2}+48v_{3}^{2}+30v_{4}^{2}-12v_{2}v_{2F}\cos(2\Psi_{2})
-177 v_2 v_{2F} v_4 \cos(2 \Psi_2 - 4 \Psi_4)+93v_{2}^{2}v_{4}\cos(4\Psi_{2} -4\Psi_{4} )\notag\\&+\frac{75}{2}v_{2F}^{2}v_{4}\cos(4\Psi _{4} )  
  + 45 v_2^2 v_{2F}^2  + 72 v_{2F}^2 v_3^2 + 45 v_{2F}^2 v_4^2 - \frac{165}{2} v_2^3 v_{2F} \cos(2 \Psi_2) - 3 v_2 v_{2F}^3 \cos(2 \Psi_2) \notag\\ &- \frac{165}{8} v_2^3 v_{2F}^3 \cos(2 \Psi_2) 
 - 246 v_2 v_{2F} v_3^2 \cos(2 \Psi_2) - \frac{123}{2} v_2 v_{2F}^3 v_3^2 \cos(2 \Psi_2) - 162 v_2 v_{2F} v_4^2 \cos(2 \Psi_2) \notag\\ &- \frac{81}{2} v_2 v_{2F}^3 v_4^2 \cos(2 \Psi_2) 
 + \frac{33}{4} v_2^2 v_{2F}^2 \cos(4 \Psi_2) - \frac{1}{8} v_2^3 v_{2F}^3 \cos(6 \Psi_2) + \frac{123}{2} v_2 v_{2F}^2 v_3^2 \cos(2 \Psi_2 - 6 \Psi_3) 
 \notag\\ &- \frac{15}{4} v_{2F}^3 v_3^2 \cos(6 \Psi_3) - \frac{23}{2} v_2 v_{2F}^3 v_4^2 \cos(2 \Psi_2 - 8 \Psi_4)
 - \frac{177}{4} v_2 v_{2F}^3 v_4 \cos(2 \Psi_2 - 4 \Psi_4)  \notag\\ &+ \frac{279}{2} v_2^2 v_{2F}^2 v_4 \cos(4 \Psi_2 - 4 \Psi_4) 
 - 120 v_{2F} v_3^2 v_4 \cos(6 \Psi_3 - 4 \Psi_4) - 30 v_{2F}^3 v_3^2 v_4 \cos(6 \Psi_3 - 4 \Psi_4) 
 \notag\\ &+ \frac{183}{2} v_2^2 v_{2F}^2 v_4 \cos(4 \Psi_4) + \frac{243}{2} v_{2F}^2 v_3^2 v_4 \cos(4 \Psi_4) + \frac{75}{2} v_{2F}^2 v_4^3 \cos(4 \Psi_4) 
 - \frac{15}{2} v_2 v_{2F}^3 v_4 \cos(2 \Psi_2 + 4 \Psi_4),\notag\\ 
D_{4} = \hspace{0.1cm}&6+322v_{2}^{2}+456v_{3}^{2}+320v_{4}^{2}-272v_{2}v_{2F}\cos(2\Psi_{2})
-2004 v_2 v_{2F} v_4 \cos(2 \Psi_2 - 4 \Psi_4)+651v_{2}^{2}v_{4}\cos(4\Psi_{2} -4\Psi_{4} )\notag\\
&+645v_{2F}^{2}v_{4}\cos(4\Psi _{4} )
+ 18 v_{2F}^2 + 966 v_2^2 v_{2F}^2 + \frac{9}{4} v_{2F}^4 + \frac{483}{4} v_2^2 v_{2F}^4  + 1368 v_{2F}^2 v_3^2 + 171 v_{2F}^4 v_3^2+ 960 v_{2F}^2 v_4^2 \notag\\ &+ 120 v_{2F}^4 v_4^2 - 914 v_2^3 v_{2F} \cos(2 \Psi_2) - 204 v_2 v_{2F}^3 \cos(2 \Psi_2)- \frac{1371}{2} v_2^3 v_{2F}^3 \cos(2 \Psi_2) \notag\\ &- 2436 v_2 v_{2F} v_3^2 \cos(2 \Psi_2) - 1827 v_2 v_{2F}^3 v_3^2 \cos(2 \Psi_2) - 1736 v_2 v_{2F} v_4^2 \cos(2 \Psi_2) 
 - 1302 v_2 v_{2F}^3 v_4^2 \cos(2 \Psi_2) \notag\\&+ \frac{441}{2} v_2^2 v_{2F}^2 \cos(4 \Psi_2) + \frac{147}{4} v_2^2 v_{2F}^4 \cos(4 \Psi_2) - \frac{15}{2} v_2^3 v_{2F}^3 \cos(6 \Psi_2) 
 + 915 v_2 v_{2F}^2 v_3^2 \cos(2 \Psi_2 - 6 \Psi_3) \notag\\ &+ \frac{305}{2} v_2 v_{2F}^4 v_3^2 \cos(2 \Psi_2 - 6 \Psi_3) - 147 v_{2F}^3 v_3^2 \cos(6 \Psi_3) 
 + \frac{9}{2} v_2 v_{2F}^4 v_3^2 \cos(2 \Psi_2 + 6 \Psi_3) - 330 v_2 v_{2F}^3 v_4^2 \cos(2 \Psi_2 - 8 \Psi_4) \notag\\&
 - 1503 v_2 v_{2F}^3 v_4 \cos(2 \Psi_2 - 4 \Psi_4)  + 1953 v_2^2 v_{2F}^2 v_4 \cos(4 \Psi_2 - 4 \Psi_4) 
 + \frac{1953}{8} v_2^2 v_{2F}^4 v_4 \cos(4 \Psi_2 - 4 \Psi_4) \notag\\ &- 1152 v_{2F} v_3^2 v_4 \cos(6 \Psi_3 - 4 \Psi_4)-864 v_{2F}^{3} v_{3}^{2} v_{4} \cos(6 \Psi_{3} - 4 \Psi_{4})  + 1548 v_{2}^{2} v_{2F}^{2} v_{4} \cos(4 \Psi_{4}) + \frac{{215}}{2} v_{2F}^{4} v_{4} \cos(4 \Psi_{4})
\notag\\&+ 258 v_2^2 v_{2F}^4 v_4 \cos(4 \Psi_4) 
 + 1776 v_{2F}^2 v_3^2 v_4 \cos(4 \Psi_4) + 296 v_{2F}^4 v_3^2 v_4 \cos(4 \Psi_4) + 603 v_{2F}^2 v_4^3 \cos(4 \Psi_4) \notag\\ &+ \frac{201}{2} v_{2F}^4 v_4^3 \cos(4 \Psi_4) 
 + \frac{119}{4} v_{2F}^4 v_4^2 \cos(8 \Psi_4) - 322 v_2 v_{2F}^3 v_4 \cos(2 \Psi_2 + 4 \Psi_4) + \frac{127}{16} v_2^2 v_{2F}^4 v_4 \cos(4 \Psi_2 + 4 \Psi_4).
\end{align}

\begin{figure}[H]
\centering
\includegraphics[scale=0.4]{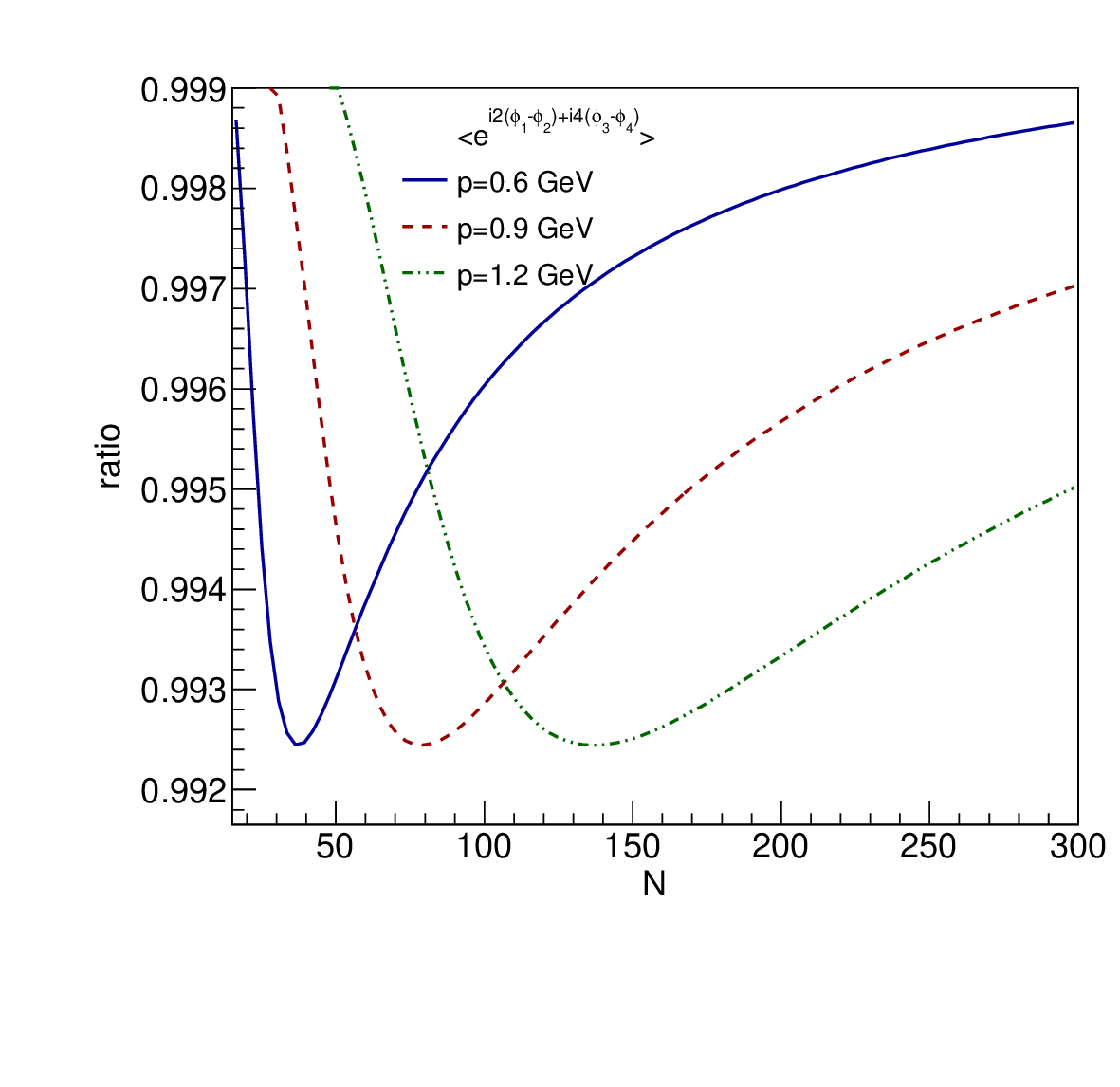}
\includegraphics[scale=0.4]{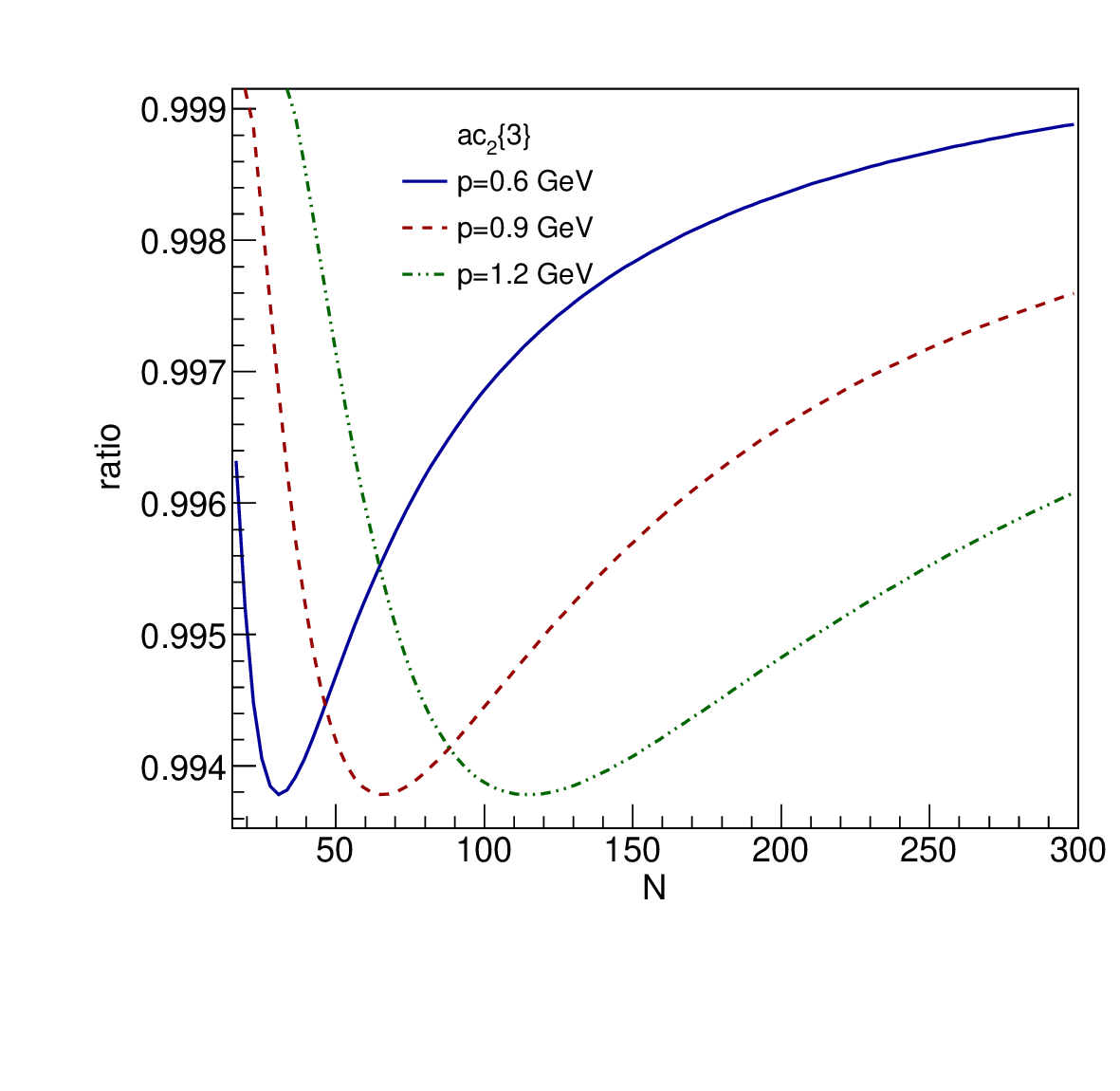}
\caption{
The ratios of the approximate result to the full result, i.e., Eq.~\eqref{eq24} / Eq.~\eqref{eqA1} (left panel) and Eq.~\eqref{eq34} / Eq.~\eqref{eqA2} (right panel), as a function of the number of particles $N$, for different values of the transverse momenta $p$}.
\label{ratio}
\end{figure}

Based on Fig.~\ref{ratio}, the ratios of the approximate results of Eqs.~\eqref{eq24} and ~\eqref{eq34} to the full results of Eqs.~\eqref{eqA1} and ~\eqref{eqA2} in this appendix are both close to 1, with the worst approximations of $99.244\%$ and $99.378\%$ respectively, suggesting that Eqs.~\eqref{eq24} and ~\eqref{eq34} can be good proxies for the results of Eqs.~\eqref{eqA1} and ~\eqref{eqA2} in this appendix.

\bibliography{ref}

\end{document}